\documentclass[aps,prb,twocolumn,superscriptaddress, showpacs]{revtex4-1}
\usepackage{amssymb,amsmath,graphicx,xspace,color}
\usepackage[pdfauthor={Christoph Wuttke}, bookmarks=true, pdfborder={0 0 0}, colorlinks=true,citecolor=blue,linkcolor=black,urlcolor=blue]{hyperref}

\newcommand{\mn}{Mn$_3$Ge\xspace}
\newcommand{\sn}{Mn$_3$Sn\xspace}

\begin{document}
\title{Berry curvature unravelled by the anomalous Nernst effect in \mn}


\author{Christoph Wuttke}
\email[]{c.wuttke@ifw-dresden.de}
\affiliation{Leibniz-Institute for Solid State and Materials Research, IFW-Dresden, 01069 Dresden, Germany}
\author{Federico Caglieris}
\affiliation{Leibniz-Institute for Solid State and Materials Research, IFW-Dresden, 01069 Dresden, Germany}
\author{Steffen Sykora}
\affiliation{Leibniz-Institute for Solid State and Materials Research, IFW-Dresden, 01069 Dresden, Germany}
\author{Francesco Scaravaggi}
\affiliation{Leibniz-Institute for Solid State and Materials Research, IFW-Dresden, 01069 Dresden, Germany}
\affiliation{Institut f\"{u}r Festk\"{o}rperphysik, TU Dresden, 01069 Dresden, Germany}
\author{Anja U. B. Wolter}
\affiliation{Leibniz-Institute for Solid State and Materials Research, IFW-Dresden, 01069 Dresden, Germany}
\author{Kaustuv Manna}
\affiliation{Max Planck Institute for Chemical Physics of Solids, 01187 Dresden, Germany}
\author{Vicky S\"{u}ss}
\affiliation{Max Planck Institute for Chemical Physics of Solids, 01187 Dresden, Germany}
\author{Chandra Shekhar}
\affiliation{Max Planck Institute for Chemical Physics of Solids, 01187 Dresden, Germany}
\author{Claudia Felser}
\affiliation{Max Planck Institute for Chemical Physics of Solids, 01187 Dresden, Germany}
\author{Bernd B\"uchner}
\affiliation{Leibniz-Institute for Solid State and Materials Research, IFW-Dresden, 01069 Dresden, Germany}
\affiliation{Institut f\"{u}r Festk\"{o}rperphysik, TU Dresden, 01069 Dresden, Germany}
\affiliation{Center for Transport and Devices, TU Dresden, 01069 Dresden, Germany}
\author{Christian Hess}
\email[]{c.hess@ifw-dresden.de}
\affiliation{Leibniz-Institute for Solid State and Materials Research, IFW-Dresden, 01069 Dresden, Germany}
\affiliation{Center for Transport and Devices, TU Dresden, 01069 Dresden, Germany}


\date{\today}


\begin{abstract}
The discovery of topological quantum materials represents a striking innovation in modern condensed matter physics with remarkable fundamental and technological implications. Their classification has been recently extended to topological Weyl semimetals, i.e., solid state systems which exhibit the elusive Weyl fermions as low-energy excitations. Here we show that the Nernst effect can be exploited as a sensitive probe for determining key parameters of the Weyl physics, applying it to the non-collinear antiferromagnet \mn. This compound exhibits anomalous thermoelectric transport due to enhanced Berry curvature from Weyl points located extremely close to the Fermi level. We establish from our data a direct measure of the Berry curvature at the Fermi level and, using a minimal model of a Weyl semimetal, extract the Weyl point energy and their distance in momentum-space.
\end{abstract}

\maketitle

\section{Introduction}

Weyl semimetals\cite{Wan2011, Yang2015, Nakatsuji2015, Armitage2018} are certainly one of the most stunning representatives of topological material classes. Their electronic band structure is predicted to host Weyl points (WP), i.e., three-dimensional linear band crossings that represent massless Weyl fermions of defined chirality. Two Weyl points always form a pair of opposite chirality which is separated in momentum space due to spin-orbit coupling and breaking of the time-reversal symmetry or inversion symmetry. WPs act as source or sink of Berry curvature, a vector field in momentum space which represents the topological properties in a material. Understanding, probing and controlling this quantity is of enormous importance to emergent fields of basic and applied research. For example, in spintronics\cite{Kimata2019} the Berry curvature is causing a spin-orbit torque that drives spin dynamics in transition-metal bilayers\cite{Kurebayashi2014}. A further example is quantum computing, where the Berry curvature provides a superior robustness to noise in photonic networks of solid-state qubits\cite{Yale2016}. 

The Berry curvature can be seen as an effective magnetic field in the reciprocal lattice, determining an additional component to the electron velocity v(\textbf{k}), the so-called anomalous velocity, which is always perpendicular to the force driving the electron motion\cite{Xiao2010}. As a natural consequence, anomalous transverse transport properties, namely the anomalous Hall effect (AHE) and its thermoelectric counterpart, the anomalous Nernst effect (ANE), are expected to arise\cite{Sundaram1999, Xiao2010, Nagaosa2010, Noky2018} and have been measured in several systems\cite{Caglieris2018, Watzman2018, Sakai2018}. The Nernst effect often is dominated by the transverse Peltier coefficient $\alpha_{ij}$, which probes the electrons only within the energy window determined by the thermal broadening of the Fermi function. The ANE is thus expected to be a more sensitive probe for the Berry curvature at the Fermi level than the AHE which probes the whole Fermi sea\cite{Li2017, Caglieris2018}.

\begin{figure}
\includegraphics[width=1\columnwidth]{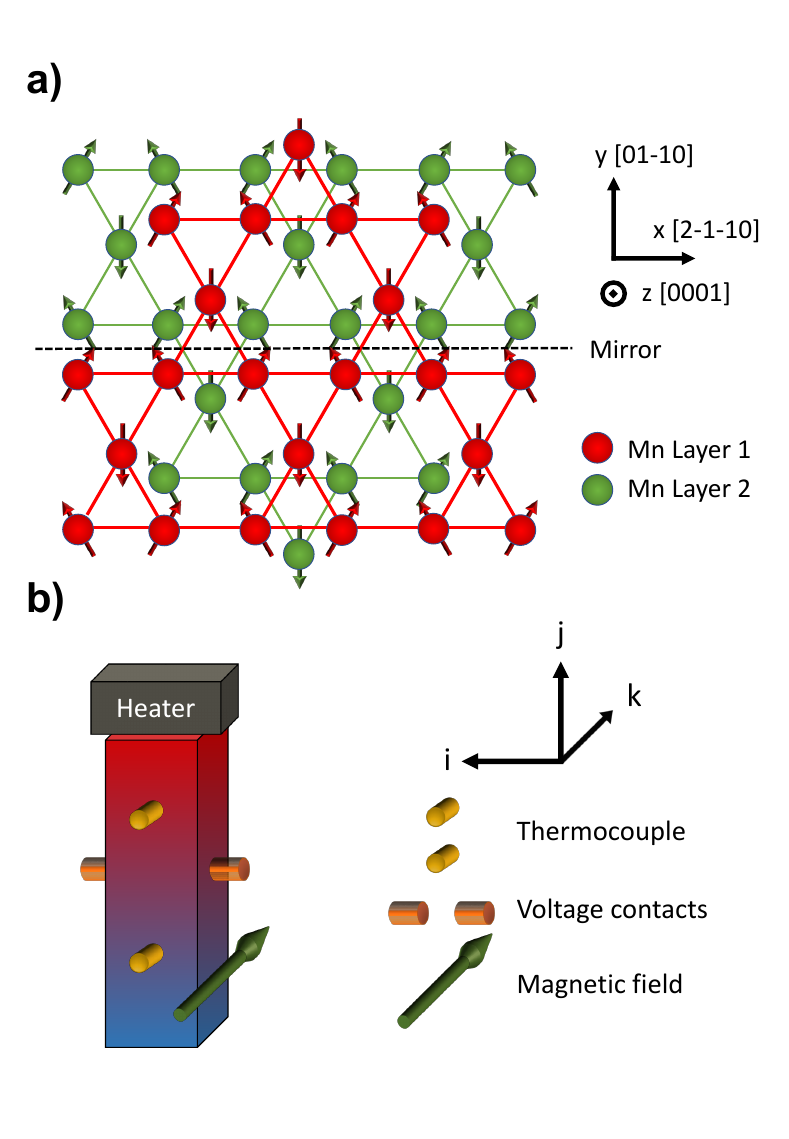}
\caption{\textbf{a)} Mn-atoms and magnetic structure of \mn. Ge atoms are in the center of each Mn-hexagon (not sketched here). By mirroring ($xz$-mirror plane) and translating by $c/2$ along the z axis the different layers are transformed into each other. \textbf{b)} Schematic Nernst setup for measuring $S_{ij}$. On the upper side a resistive chip heater creates the thermal gradient. The bottom of the sample is coupled to a thermal bath of controlled temperature, therefore the temperature gradient arises along $j$-direction. The magnetic field is applied in $k$-direction and the Nernst signal is measured along $i$-direction.}
\label{fig:Struct_meas}
\end{figure}

Recently, the isostructural non-collinear antiferromagnets \mn and \sn attracted a tremendous interest\cite{Kimata2019} due to giant anomalous transport coefficients\cite{Nayak2016, Nakatsuji2015, Ikhlas2017, Li2017, Chen2014, Guo2017}. \sn has been extensively studied in Hall-\cite{Nakatsuji2015} and Nernst-measurements\cite{Ikhlas2017, Li2017}, but it lacks magnetic order below $T = 50~$K and develops a glassy ferromagnetic ground state\cite{Brown1990, Feng2006}, where both ANE and AHE vanish\cite{Li2017}. This is in contrast to \mn, where magnetic order and anomalous transport persist down to lowest temperatures, and which is at the focus of this paper. \mn is characterized by a hexagonal crystal structure (space group $P6_3/mmc$), where Mn atoms form a kagome lattice of mixed triangles and hexagons with Ge atoms being situated at the center of the hexagons. In the noncollinear antiferromagnetic ground state of \mn ($T_N \approx~365$ to $400~$K\cite{Qian2014, Yamada1988, Tomiyoshi1982, Nagamiya1982}) the Mn moments are oriented at 120$^\circ$ with respect to their neighbors\cite{Yang2017} [see Fig.~\ref{fig:Struct_meas}(a)]. Only a very small net moment of $\approx 0.02~\mu_{B}$ appears in-plane due to a slight tilting of the Mn-moments\cite{Yamada1988, Tomiyoshi1983}. Multiple Weyl points have been predicted to exist in the band structure of \mn\cite{Yang2017}. 

Here we report a comprehensive study of the ANE\footnote{\label{Preprint}Upon finalizing this manuscript we became aware of the research of Xu \textit{et al.} (arXiv:1812.04339). The Nernst data is in well agreement to our measurement, however, only one configuration of $S_{ij}$ was measured and the focus is lying on the relations between different transverse transport coefficients such as the Wiedemann-Franz law. The calculated SOC-induced gap that this work is showing does not influence our argument.} in the noncollinear antiferromagnet \mn. We observe at all temperatures studied that the Nernst effect is dominated by a field-saturated anomalous contribution if an in-plane magnetic field $B > 0.02$~T is applied. We derive the anomalous transverse Peltier coefficient from the ANE data and show that its temperature dependence can be analyzed to extract key properties of the Weyl semimetal, i.e. the Weyl point energy, the momentum space separation of two Weyl points, and the effective strength of the Berry curvature at the Fermi level.

In Sec.~\ref{sec:experimental} we describe the experimental details. In Sec.~\ref{sec:Results_data} we show corresponding experimental results and demonstrate that the Nernst effect in \mn is purely anomalous. Our theoretical model to analyze the temperature dependence of the anomalous Peltier and Hall coefficients is presented in Sec.~\ref{sec:theory} and furthermore applied to \mn where important Weyl point properties are extracted from experimental data. Finally, we discuss and summarize our results in Sec.~\ref{sec:discussion} and Sec.~\ref{sec:summary}.

\section{Experimental}
\label{sec:experimental}

The \mn single crystals were grown using the Bridgman-Stockbarger technique. First, the high purity metals were premelted in an alumina crucible using induction melting. Then the crushed powder was filled in a custom-designed sharp-edged alumina tube and sealed inside a tantalum tube. The crystal growth temperature was controlled using a thermometer at the bottom of the ampule. The sample was heated to $1000 ^\circ\text{C}$, held there for 12~h to ensure homogeneous mixing of the melt, and then slowly cooled to $750 ^\circ\text{C}$. Finally, the sample was quenched to room temperature to retain its high temperature hexagonal phase. The single crystallinity was checked by white-beam backscattering Laue x-ray diffraction at room temperature. The crystal structures were analyzed with a Bruker D8 VENTURE x-ray diffractometer using Mo-K radiation.

Thermoelectric measurements were done in a home-built probe in a Helium cryostat with a magnetic field of up to 15~T. The thermal gradient is generated with a chip resistor on one end of the sample, the other end is glued to a cold bath with an Al$_2$O$_3$-plate in between to establish electrical current free conditions. The gradient along the sample is measured with a magnetic field calibrated AuFe/Chromel-P thermocouple. The Nernst voltage is measured perpendicular to the thermal gradient and the applied field. The different measurement configurations are labeled as $S_{ij}$, with the Nernst signal measured along the $i$-direction with an applied temperature gradient $\nabla T_j$ and a magnetic field $B_k$ [compare Fig.~\ref{fig:Struct_meas}(b)].

Due to its general dependence on the temperature as well as the magnetic field, the Nernst signal is usually measured in two different modes. A temperature dependent Nernst signal measurement contains two separate temperature sweeps. During the first sweep the magnetic field is fixed to a certain field, the second sweep is measured at the corresponding inverted magnetic field (in this work $B = 14$~T and $B = -14$~T). Afterwards the data is antisymmetrized to get rid of any contribution of the Seebeck effect caused by slightly misaligned contacts. A magnetic field dependent measurement is conducted at a fixed temperature while sweeping the magnetic field from negative to positive values or vice versa. This was not possible for the magnetic field dependent measurements in \mn, due to the peculiar hysteretic behaviour at small fields, which leads to different $S_{ij}(B)$ curves depending of the field history. Therefore, the Nernst signal was measured in full field cycles, from $B = -15$~T to $+15$~T and back to $-15$~T. By subtraction of the symmetric contribution (which shows no field dependence), the curves were centered around $S_{ij}(B)=0$ to allow a comparison of different temperatures.

\section{Nernst effect results}
\label{sec:Results_data}

\begin{figure}
\includegraphics[width=1\columnwidth]{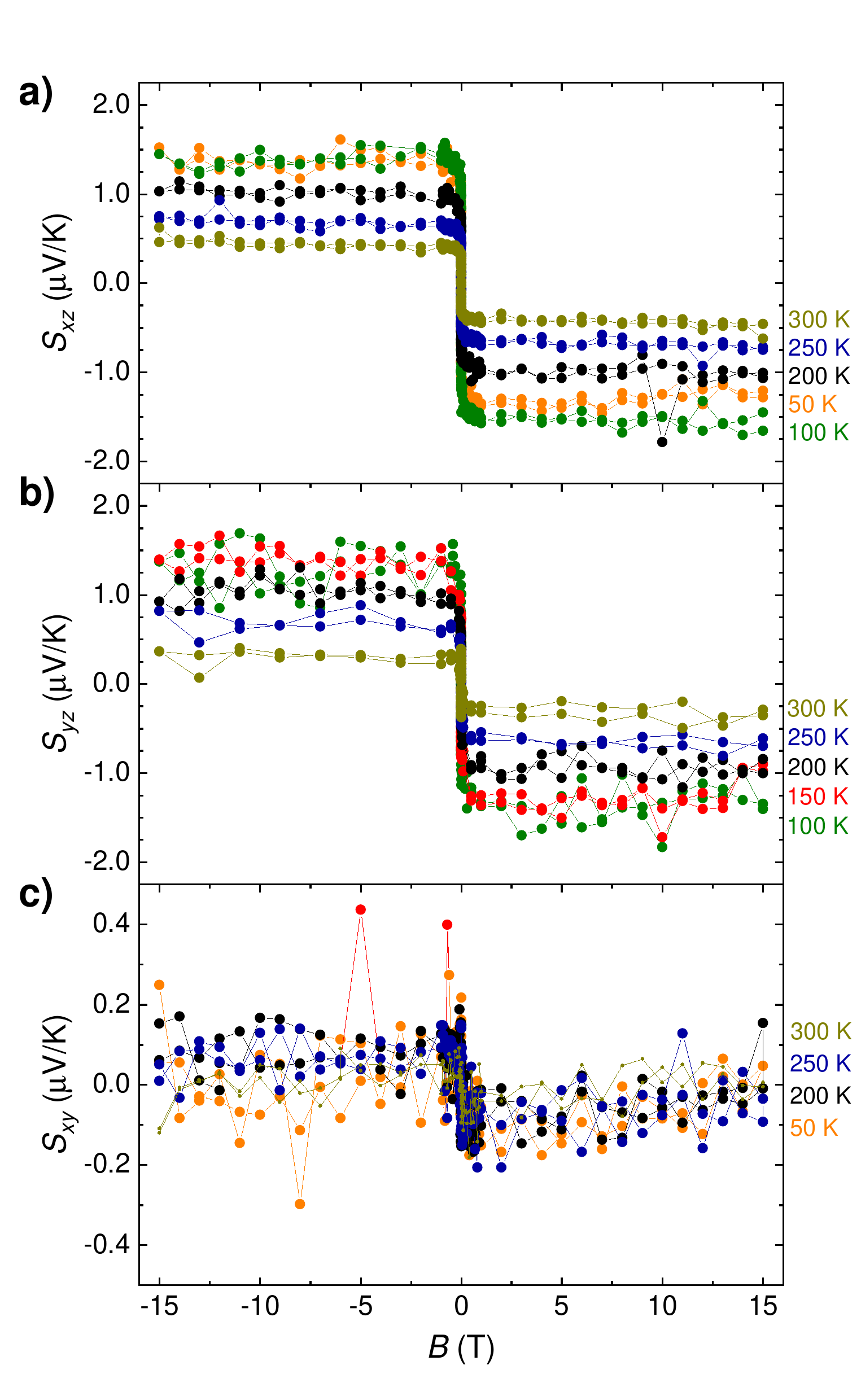}
\caption{\textbf{a), b), c)} Nernst signal of the \mn single crystals with respect to the applied magnetic field $B$ in different configurations. $S_{ij}$ is obtained by measuring the voltage along the $i$-direction while applying a thermal gradient $\nabla T$ and a magnetic field $B$ along $j$ and $k$-direction, respectively.}
\label{fig:S_ij(B)}
\end{figure}

We start with a clear demonstration that the anomalous transport (which is driven by Berry curvature) dominates the Nernst effect in \mn. Figure \ref{fig:S_ij(B)}(a) shows the Nernst coefficient $S_{xz}$ which exhibits a totally anomalous behavior (no magnetic field dependence) without any visible normal (linear $B$-dependence) contribution as a function of field for all the investigated temperatures, exhibiting a step-like feature at very low fields and reaching a saturation in a flat plateau for $B > 0.02~$T (the distinct step-like behavior at $B < 0.02~$T as well as the relation between Nernst data and magnetization are discussed in detail in appendix~\ref{sec:magn}). $S_{yz}$ reveals a very similar field dependence as presented in Fig.~\ref{fig:S_ij(B)}(b). Both configurations show the peculiar saturating behavior up to room temperature with a large Nernst signal of around 0.4 - 1.5~$\mu$V/K, depending on the temperature. On the other hand, a different phenomenology characterizes $S_{xy}$, as reported in Fig.~\ref{fig:S_ij(B)}(c). In this configuration the Nernst coefficient is much smaller, with the step-like behavior just slightly visible, and shows a much weaker temperature dependence. 

The experimental observation in Fig.~\ref{fig:S_ij(B)}(a) allows us to draw an important conclusion about the Berry curvature in \mn. The Nernst signal $S_{xz}$ of the transverse transport is generally determined by the thermoelectric tensor $\alpha$ and the charge conductivity tensor $\sigma$ as
\begin{equation}
\label{S_xz}
S_{xz} = \frac{\alpha_{xz} \sigma_{xx} - \alpha_{xx} \sigma_{xz}}{\sigma_{xx}^2 + \sigma_{xz}^2},
\end{equation}
where each transverse transport coefficient ($S_{xz}$, $\alpha_{xz}$, $\sigma_{xz}$) is the sum of a normal and an anomalous contribution. The observed saturation in the Nernst signal without \textit{any} field dependence at $B > 0.02$~T is incompatible with normal transport\cite{Caglieris2018, Ikhlas2017} and thus implies that all normal contributions are negligible in $S_{xz}$. Hence, in what follows we solely consider only anomalous transport coefficients $\alpha_{xz}$ and $\sigma_{xz}$ for analyzing our data. It is well established that the anomalous Peltier and Hall coefficients $\alpha_{xz}$ and $\sigma_{xz}$ are related to the $y$ component of the momentum integrated Berry curvature ${\bf \Omega} = (\Omega^{x}, \Omega^{y}, \Omega^{x})$ via the expressions
\begin{eqnarray}
\label{sigma}
\sigma_{xz} &=&  \frac{e^2}{\hbar}  \int \frac{d^3k}{(2\pi)^3} \Omega^{y}({\bf k}) f_{\bf k}, \\
\alpha_{xz} &=&  \frac{k_Be}{\hbar} \int \frac{d^3k}{(2\pi)^3} \Omega^{y}({\bf k}) s_{\bf k},
\label{alpha_main}
\end{eqnarray}
where $f_{\bf k}$ is the Fermi distribution function and $s_{\bf k}$ is the entropy density\footnote{For simplicity, a special index for highlighting those coefficients as anomalous is omitted in the following.}. Thus, if a large anomalous Nernst signal $S_{xz}$ is observed, the $y$ component of the integrated Berry curvature must be large too. The almost identical observation for $S_{yz}$ [see Fig.~\ref{fig:S_ij(B)}(b)] implies the same statement for $\Omega^{x}$. Since $S_{xy}$ exhibits only a suppressed value without any clear anomalous contribution, these considerations allow us to conclude that the integrations of the $x$- and $y$-components of the Berry curvature are large compared to its integrated $z$ component, caused by an either vanishing or odd $\Omega^{z}$ with respect to $k_{z} = 0$ ($\Omega^{z}(k_{x}, k_{y}, k_{z}) = -\Omega^{z}(k_{x}, k_{y}, - k_{z})$). These findings are consistent with symmetry considerations of the band structure\cite{Yang2017, Tomiyoshi1982}, according to which $\Omega^{z}$ is an odd function in $k_{z}$, whereas $\Omega^{x}$ is even in $k_{x}$ if the magnetic field is applied along $x$ and likewise for $\Omega^{y}$.  

\begin{figure}
\includegraphics[width=1\columnwidth]{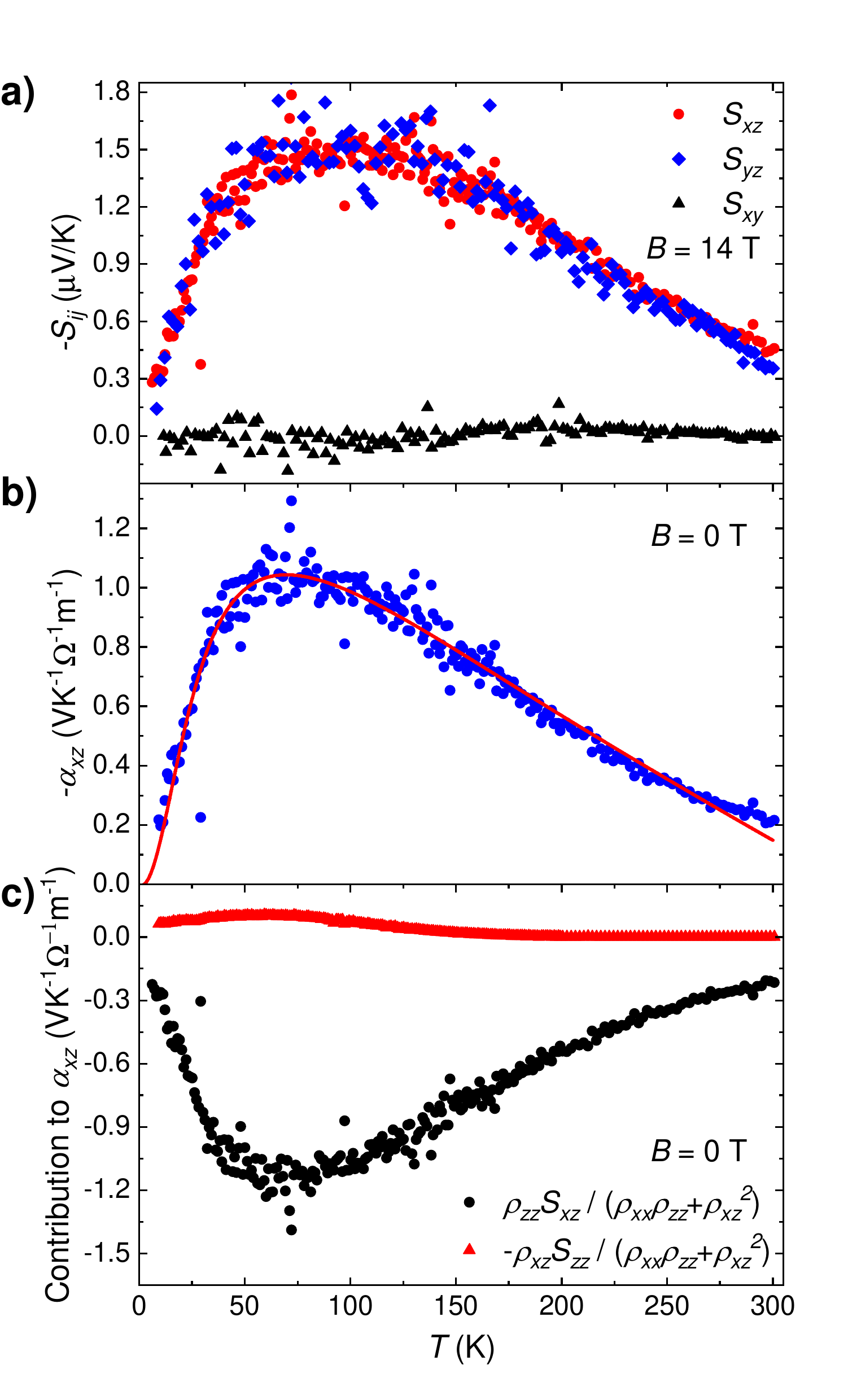}
\caption{Temperature dependence of \textbf{a)} $S_{ij}$ at $B = 14$~T, \textbf{b)} $\alpha_{xz}$ obtained by using the relation $\alpha_{xz} = \left(\rho_{zz}S_{xz} - \rho_{xz}S_{zz}\right)/\left(\rho_{xx}\rho_{zz} + \rho_{xz}^{2}\right)$\cite{Meinero2018} (blue dots) and the fit provided by our theoretical model (red line), \textbf{c)} the two terms contributing to $\alpha_{xz}$.}
\label{fig:S_ij(T)}
\end{figure}

The saturation values of $S_{xz}$, $S_{yz}$ and $S_{xy}$ at $B = 14$~T are plotted in Fig.~\ref{fig:S_ij(T)}(a) as a function of temperature $T$. A broad maximum of about 1.5~$\mu$V/K is visible for $S_{xz}$ and $S_{yz}$ at around 100~K, whereas $S_{xy}$ is negligibly small. This remarkable temperature dependence is leading us to a second qualitative fundamental conclusion. As is explained in Ref.~\onlinecite{Caglieris2018}, the peak position in the temperature dependence of the ANE represents a coarse correspondence with the lowest Weyl point energy $\mu$ with respect to the Fermi level. This is because for $k_BT\ll|\mu|$ essentially states with energy $|\epsilon|<|\mu|$ probe the Berry curvature and contribute to the ANE, giving rise to an increase of it upon the thermal energy approaching $|\mu|$ from below. On the other hand, if $k_BT$ becomes comparable with $|\mu|$ or even exceeds it, the then additionally contributing higher energy states at $|\epsilon| > |\mu|$ provide an opposite contribution to the ANE. Thus, we estimate $|\mu|\sim10$~meV. It has been suggested that in \mn the presence of spin-orbit coupling removes the degeneracy at the Weyl point and leads to an opening of a gap\cite{Xu2019}. In this case the energy $|\mu|$ remains meaningful and describes the distance between the center of the gap and the Fermi level.

After having established a qualitative understanding of the ANE, we now move on to extract material specific parameters of the Weyl system. In order to provide a quantitative evaluation we derived the transverse Peltier coefficient 
\begin{equation}
\label{Alpha_xz_calculation}
\alpha_{xz} = \frac{\rho_{zz}S_{xz} - \rho_{xz}S_{zz}}{\rho_{xx}\rho_{zz} + \rho_{xz}^{2}}
\end{equation}
using experimental values for the required transport coefficients [compare Appendix~\ref{sec:peltier}]. The temperature dependence of $\alpha_{xz}$ in zero field is shown in Fig.~\ref{fig:S_ij(T)}(b). It resembles the behaviour of $S_{xz}$ but exhibits a narrower maximum which is shifted to lower temperatures ($\approx$~75~K). It is worth marking that indeed the anomalous Nernst signal dominates $\alpha_{xz}$. This can be inferred from Fig.~\ref{fig:S_ij(T)}(c) where the two contributions $\rho_{zz}S_{xz} / (\rho_{xx}\rho_{zz} + \rho_{xz}^{2})$ and $- \rho_{xz}S_{zz} / (\rho_{xx}\rho_{zz} + \rho_{xz}^{2})$ are directly compared. Thus $\alpha_{xz}$ is truly anomalous as well and is given by Eq.~\eqref{alpha_main}. We use this as a starting point to develop a model to analyze the temperature dependence of $\alpha_{xz}$ in more detail and to extract important Weyl point properties from our experimental result of Fig.~\ref{fig:S_ij(T)}(b).

\section{Theoretical modelling and data analysis}
\subsection*{Model}
\label{sec:theory}

\begin{figure}
\includegraphics[width=1\columnwidth]{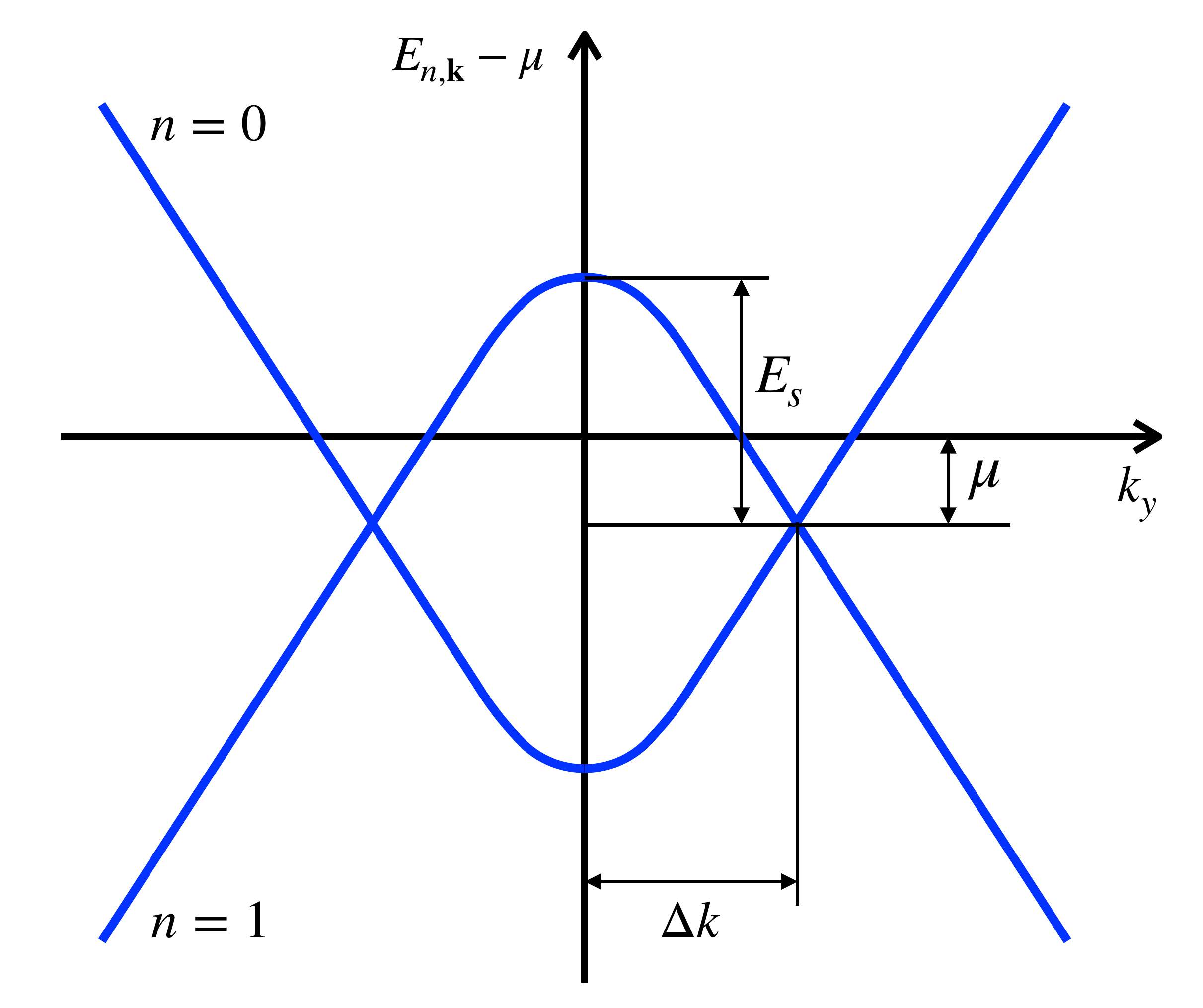}
\caption{Schematic picture of the band dispersion in a Weyl semimetal. The Weyl points of opposite chirality are placed along the $k_y$ direction. Saddle point energy, energy of the Weyl points with respect to the Fermi level and their distance from $k_y = 0$ are labeled as $E_S$, $\mu$ and $\Delta k$ respectively.}
\label{fig:model}
\end{figure}

Extending the theory proposed in Ref.~\onlinecite{Caglieris2018} we start from a generalized expression of Eq.~\eqref{alpha_main} for a multiband system, 
\begin{eqnarray}
\alpha_{xz} &=&  \frac{e^2}{\hbar} \sum_n \int \frac{d^3k}{(2\pi)^3} \Omega_n^{y}({\bf k}) s_{n,{\bf k}},
\label{alpha}
\end{eqnarray}
where $s_{n,{\bf k}} = - f_{n,{\bf k}} \ln f_{n,{\bf k}} - \left[ (1 - f_{n,{\bf k}}) \ln (1 - f_{n,{\bf k}}) \right]$ is the entropy density ($f_{n,{\bf k}} = f(E_{n{\bf k}})$: Fermi distribution function) for the dispersion $E_{n,{\bf k}}$ of the conduction electron band $n$. We assume that this anomalous contribution is predominantly determined by one particular pair of Weyl nodes which are placed extremely close to the Fermi level\cite{Yang2017}. The vector $\mathbf{\Omega}_n({\bf k})$ is the Berry curvature with respect to the band '$n$'. We consider two bands which are separated in energy and which touch each other in the pair of Weyl nodes [compare Fig.~\ref{fig:model}]. They are indexed by $n=0$ (high energy band with $E_{0,{\bf k}} > 0$) and $n=1$ (low energy band with $E_{1,{\bf k}} < 0$). For the dispersion we apply the following minimal model\cite{Sharma2016} of linearized Weyl fermions,
\begin{eqnarray}
E_{n,{\bf k}}=\pm v_F \sqrt{k_x^2 + (k_y \pm \Delta k)^2 + k_z^2}.
\label{dispersion}
\end{eqnarray}
It describes a pair of Weyl points which are placed at energy $E=0$ and are separated in momentum space by $2\Delta k$.

We replace at first the momentum integral in Eq.~\eqref{alpha} with an energy integral,
\begin{eqnarray}
\alpha_{xz} &=&  \frac{e^2}{\hbar}  \int_{-\infty}^{0} dE \, \rho_1(E) \Omega_1^{y}(E) s(E) \nonumber\\ 
&+&  \frac{e^2}{\hbar} \int_{0}^{\infty} dE \, \rho_0(E) \Omega_0^{y}(E) s(E),
\label{alpha_energy_1}
\end{eqnarray} 
where $\rho_n(E)$ is the density of states with respect to the band '$n$'. For the linearized band structure \eqref{dispersion} the density of states is given by
\begin{eqnarray}
\label{rho_0}
\rho_0(E) &=& \left\{
\begin{array}{rl}
\displaystyle 0 & :\quad E < 0 \\
\displaystyle  \rho_0 E^2 & :\quad 0 \le E \le E_s \\
\displaystyle \frac{\rho_0}{2} E (E_s + E) & :\quad E > E_s
\end{array}
\right., \\
\label{rho_1}
\rho_1(E) &=& \left\{
\begin{array}{rl}
\displaystyle 0 & :\quad E > 0 \\
\displaystyle  \rho_0 E^2 & :\quad -E_s \le E \le 0 \\
\displaystyle \frac{\rho_0}{2} E (-E_s + E) & :\quad E < -E_s
\end{array}
\right.,
\end{eqnarray}
where $\rho_0$ is a constant in energy and $E_s = v_F \Delta k$ is the energy difference between the Weyl point and the saddle point at ${\bf k} = 0$ [compare Fig.~\ref{fig:model}]. Note that the total density of states $\rho = \rho_0 + \rho_1$ is symmetrical around the Weyl point, i.~e.~$\rho(-E) = \rho(E)$, and it vanishes at the Weyl points $E=0$. The explicit formula for the energy dependence of the entropy density $s(E)$ in Eq.~\eqref{alpha_energy_1} reads
\begin{eqnarray}
\label{entropydens}
s(E) = \frac{\ln \left(1 + e^{\beta (E - \mu)} \right)}{1 + e^{\beta (E - \mu)}} + \frac{\ln \left(1 + e^{-\beta (E - \mu)} \right)}{1 + e^{-\beta (E - \mu)}},
\end{eqnarray} 
where $\mu$ is the chemical potential and $\beta = 1 / (k_BT)$ the inverse temperature. Note that the chemical potential might also be temperature-dependent. 

The vector of the Berry curvature in Eqs.~\eqref{alpha} and \eqref{alpha_energy_1} can be obtained by a standard linear response theory as introduced by Kubo\cite{Kubo1957}. The general formula for the vector of the Berry curvature in a multiband system reads 
\begin{eqnarray}
 \mathbf{\Omega}_n({\bf k}) = i \hbar^2 \sum_{m \ne n} \frac{\langle \psi_{n,{\bf k}} | \hat{\bf v} | \psi_{m,{\bf k}} \rangle \times \langle \psi_{m,{\bf k}} | \hat{\bf v} | \psi_{n,{\bf k}} \rangle}{(E_{n,{\bf k}} - E_{m,{\bf k}})^2}, 
 \label{Omega}
\end{eqnarray}
where $|\psi_{n,{\bf k}}\rangle$ are the Bloch states and $\hat{\bf v}$ is the velocity operator. Within our linearized model we neglect the energy dependence of the velocity matrix elements and assume that the Berry curvature is equal for the two bands. Thus, the energy dependence of one component of $ \mathbf{\Omega}_n$ is determined by the quadratic energy denominator in the Kubo formula \eqref{Omega}. Hence, we set for the $y$-component
\begin{eqnarray}
\label{Omega_y}
\Omega_0^y(E) = \Omega_1^y(E) = \frac{\hbar^2 \tilde{\Omega}}{4E^2},
\end{eqnarray}
with an open parameter $\tilde{\Omega}$ representing the off-diagonal velocity matrix elements in Eq.~\eqref{Omega} in a field which breaks the time-reversal symmetry of the system. Note that the Berry curvature diverges at the Weyl-point energy $E=0$. 

Inserting the expressions \eqref{rho_0}, \eqref{rho_1}, and \eqref{Omega_y} in the formula \eqref{alpha_energy_1} for $\alpha_{xz}$ we obtain
\begin{eqnarray}
\alpha_{xz} &=& \frac{e^2\hbar \tilde{\Omega} \rho_0}{4} \left[ \int_{-E_s}^{E_s} s(E) \, dE \right. \nonumber \\
&&+  \frac{1}{2}  \int_{-\infty}^{-E_s}  \frac{-E_s + E}{E} s(E) \, dE \nonumber \\
&&+ \left. \frac{1}{2}  \int_{E_s}^{\infty}  \frac{E_s + E}{E} s(E) \, dE \right],
\end{eqnarray}
where for $s(E)$ the expression \eqref{entropydens} (considering $\mu$ and $\beta$ as constant in energy) has to be used. To simplify the solution of the integrals with the aim of obtaining a fitting formula for the temperature dependence we substitute at first the variable $E$ with the Fermi distribution function $F(E) = 1 / (1 + e^{\beta E})$. Exploiting the property that $F(E)$ can only take values between 0 and 1 we then perform a Taylor expansion in terms of F around the value $1/2$ up to the first order in $(F - 1/2)$. Due to the exponential functions in $s(E)$ the power series of $F$ converges quickly if the temperature is not too large. After this expansion we integrate over $F$ and obtain the following result
\begin{eqnarray}
\label{Fitting_formula}
\alpha_{xz} &=& C_\alpha k_B T \left[ 1 - \frac{E_s}{\mu} -  \frac{E_s k_B T}{\mu^2} \right.  \\
&+& F(-E_s - \mu) \left(1 + \frac{E_s}{\mu} -  2\frac{E_s k_B T}{\mu^2} \right)  \nonumber \\
 &+&  F(E_s - \mu) \left(-1 + \frac{E_s}{\mu} -  2\frac{E_s k_B T}{\mu^2} \right)  \nonumber \\
 &+& \left. 2\frac{E_s k_B T}{\mu^2} \left( F^2(E_s - \mu) - F^2(-E_s - \mu) \right) \right], \nonumber
\end{eqnarray}
with $F(E) = 1 / (1 + e^{\beta E})$ and $C_\alpha = (e^2\hbar \tilde{\Omega} \rho_0  \ln 2)/2$. Note that the chemical potential $\mu$, which is derived below, is in general temperature-dependent, i.~e.~$\mu = \mu(T)$. After inserting this function $\mu(T)$ into Eq.~\eqref{Fitting_formula} we obtain the desired fitting formula for the temperature dependence of the Peltier coefficient $\alpha_{xz}$. 

As can be seen in Eq.~\eqref{Fitting_formula}, the parameter $C_\alpha$ plays a specific role which is used in our analysis. $C_\alpha = (e^2\hbar \tilde{\Omega} \rho_0  \ln 2)/2$, where $\tilde{\Omega}$ is according to Eqs.~\eqref{Omega} and \eqref{Omega_y} a parameter which characterizes the off-diagonal velocity matrix elements of the Berry curvature. $\rho_0$ is defined in Eqs.~\eqref{rho_0} and \eqref{rho_1} and represents the amplitude of the density of states which is mainly determined by the band dispersion. 

Let us finally derive an approximate expression for $\mu(T)$. Generally, this function can be found from the relation between the total particle number and the chemical potential which is given by an integral over the Fermi distribution as follows,
\begin{eqnarray}
\label{Relation_N_mu}
N = \int_{-\infty}^{\infty} F(E - \mu) \rho(E) dE,
\end{eqnarray}
where $\rho(E) = \rho_0(E) + \rho_1(E)$  is the total density of states with the two parts $\rho_0$ and $\rho_1$ as given by the Eqs.~\eqref{rho_0} and \eqref{rho_1}. Evaluating the energy integral in Eq.~\eqref{Relation_N_mu} and then solving the resulting expression for $\mu$ gives rise to the function $\mu = \mu(N,T)$. Unfortunately, the exact solution can only be found numerically. To find an approximate analytical formula for $\mu(T)$ we simplify the integration in Eq.~\eqref{Relation_N_mu} by replacing the exponential behavior of the Fermi distribution function $F(E-\mu)$ around $E = \mu$ with a linear function in $E$. More specifically, we model the function $F(E-\mu)$ to linearly drop to zero in the energy range $k_BT$ around $E=\mu$. For $E$ values below and above this range we set $F$ equal to 1 and 0, respectively. Such an approximation is valid if the temperature is not loo large ($k_BT < 2E_s$). We obtain from  Eq.~\eqref{Relation_N_mu}
\begin{eqnarray}
\label{Relation_N_mu_appr}
N &\approx& \int_{-\infty}^{-E_s} \frac{\rho_0}{2} E (-E_s + E) dE \nonumber \\
&+& \int_{-E_s}^{\mu - \frac{k_BT}{2}} \rho_0 E^2 dE +  \rho_0 \mu^2 \frac{k_BT}{2}.
\end{eqnarray}
Forming the derivative with respect to $(k_BT)$ on both sides of Eq.~\ref{Relation_N_mu_appr} leads to the following approximate differential equation for the chemical potential,
\begin{eqnarray}
\label{Diff_eq_mu}
0 \approx \mu^2 \mu' + \mu \frac{k_B T}{2} + \left( \frac{k_B T}{2} \right)^2 \left(  \mu' -  \frac{1}{2} \right).
\end{eqnarray}
At zero temperature $T=0$ we immediately find $\mu'=0$. In the high-temperature limit $\mu \ll k_BT$, Eq.~\eqref{Diff_eq_mu} suggests a linear behavior of $\mu$ with temperature, i.~e.~$\mu \propto k_BT$. Therefore, we assume the following approximate temperature behavior,
\begin{eqnarray}
\label{ansatz_mu_T}
 \mu (T) \approx \sqrt{ \mu_0^2 + (A k_BT)^2},
\end{eqnarray}
where $\mu_0$ is the chemical potential at zero-temperature, i.~e.~$\mu_0 = \mu(T=0)$. This ansatz fulfills the above properties as one can easily verify by considering the corresponding limiting cases. According to Fig.~\ref{fig:model}, $\mu_0$ defines the energy of the Weyl point relative to the Fermi level. The dimensionless constant $A$ can be obtained by inserting the ansatz \eqref{ansatz_mu_T} into the differential equation \eqref{Diff_eq_mu} in the limit $k_BT \ll \mu$. We find the following equation for $A$,
\begin{eqnarray}
A^3 + \frac{3}{4} A - \frac{1}{8} \approx 0,
\end{eqnarray}
which has the solution $A \approx 0.162$. Note that this constant determines the change of the chemical potential with temperature in the high-energy regime. Therefore, in the real multiband material where several  Weyl points are present we expect for $A$ a value larger than 0.162. Qualitatively, $A$ is a measure for degrees of freedom of the electronic system at the Fermi level. In the case of trivial bands being absent it roughly corresponds to the number of high-energy Weyl points in the material times the number 0.162. Therefore we can slightly modify Eq.~\eqref{ansatz_mu_T}, using the quantity of $N_W$ instead of $A$,
\begin{eqnarray}
\label{mu_T_with_N_W}
 \mu (T) \approx \sqrt{ \mu_0^2 + (0.162\cdot N_{W} \cdot k_BT)^2}.
\end{eqnarray}

In summary, the expression \eqref{Fitting_formula} together with Eq.~\eqref{ansatz_mu_T} provides a fitting formula  for the temperature dependence of the Peltier coefficient $\alpha_{xz}$. The fitting parameters are $A$, $C_\alpha$, $\mu_0$, and $E_S$. The saddle point energy $E_S = v_F \Delta k$ is related to the separation of the Weyl points of opposite chirality in momentum space and $\mu_0$ describes the energy difference between the Weyl point and the Fermi level at $T=0$. Note that $\mu_0$ is usually considered in band structure calculations as the Weyl point energy and is therefore of particular interest. Within our model the values $E_s$ and $\mu_0$ are assigned to the particular low-energy Weyl point which arises from the crossing of our two linearized bands. 
These energy values are related to the Weyl point with the lowest possible energy.

The remaining parameter $C_\alpha = (e^2\hbar \tilde{\Omega} \rho_0  \ln 2)/2$ is a constant which represents the order of magnitude of $\alpha_{xz}$, directly proportional to the amplitude of the density of states $\rho_0$ and the experimentally relevant strength of the Berry curvature of the considered Weyl system near the Fermi level, $\tilde{\Omega}$.

\subsection*{Analysis of \mn data}
\label{sec:analysis}

In order to extract the important geometric properties of the underlying system of Weyl fermions close to the Fermi level, we fit the experimental data [red line in Fig.~\ref{fig:S_ij(T)}(b)] using the formula \eqref{Fitting_formula}. As can be seen clearly, the fit works well in a wide temperature range. The deviation at high temperature above 250~K can be explained within the approximation made to derive Eq.~\eqref{Fitting_formula}. The obtained parameter $\mu_0 = 6.6 \pm 0.7$~meV is remarkably close to 8~meV, the energy of the particular Weyl point $W_4$ provided by band structure calculations\cite{Yang2017} of \mn. From the saddle point energy $E_s = 90 \pm 25$~meV and $v_F\approx$ 1 eV/$\pi$\cite{Yang2017} we calculate a momentum space separation $\Delta k \approx 0.09~\pi$ using $E_s = v_F \Delta k$. Both, this result and estimated total number of Weyl points $N_W = 17.8 \pm 2.2$ also agree well with band structure calculations\cite{Yang2017}. Furthermore, while the above parameters determine the momentum space properties, the parameter $C_\alpha = 0.030 \pm 0.002~\mathrm{V}\cdot\left(\mathrm{K~\Omega~m}\right)^{-1}$ contains the materials specific information on the Berry curvature at the Fermi level.

We mention, that the anomalous Hall coefficient can be analyzed using a similar approach. However, since an additional approximation is needed for deriving an analytical expression for the AHE, the accuracy of such an analysis of $\sigma_{xz}$ is comparably lower (see appendix~\ref{sec:Hall_fit} for details).

\section{Discussion}
\label{sec:discussion}

\begin{figure}
\includegraphics[width=1\columnwidth]{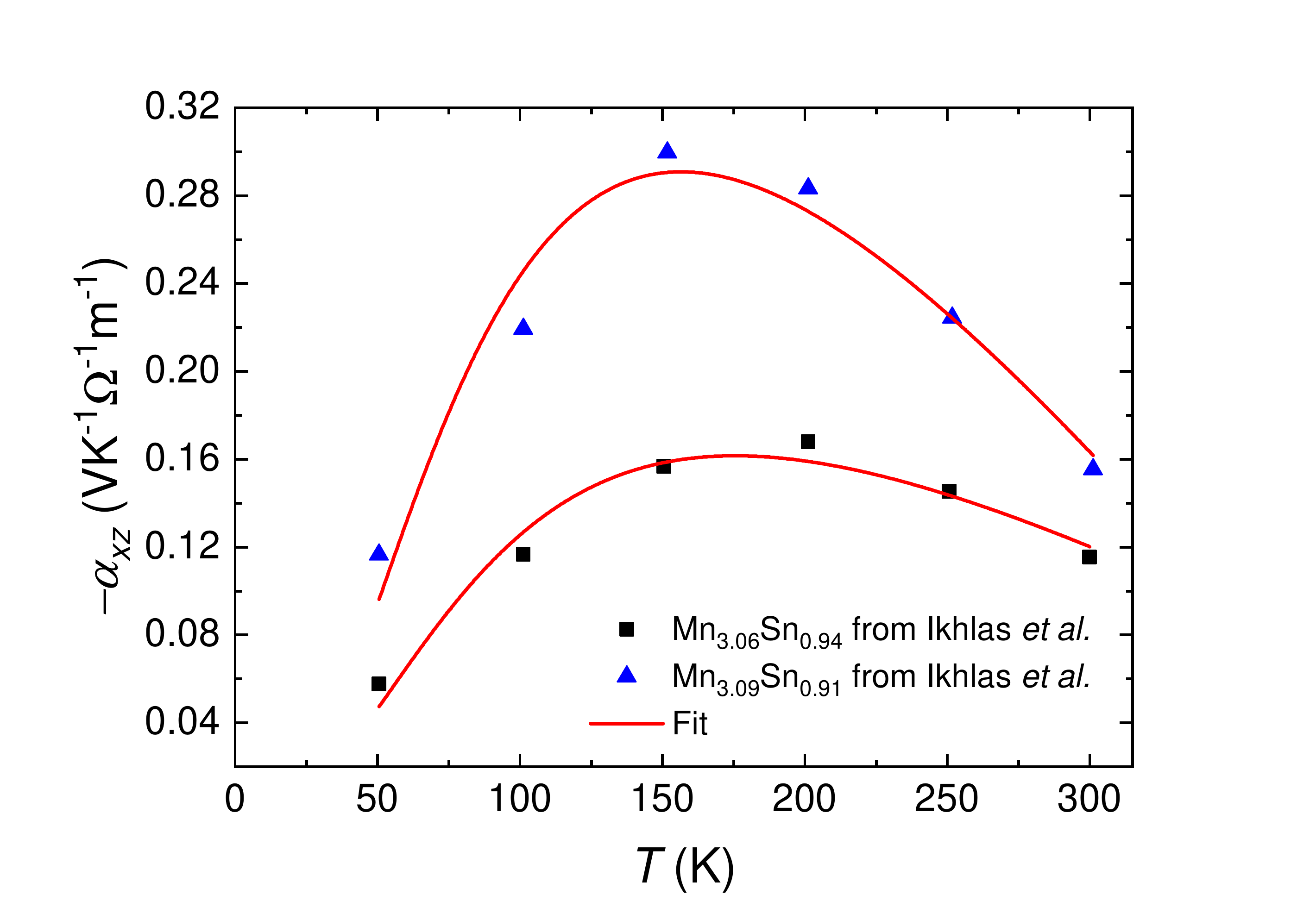}
\caption{$\alpha_{xz}$ in Mn$_{3.06}$Sn$_{0.94}$ (black squares) and Mn$_{3.09}$Sn$_{0.91}$ (blue triangles) from Ref.~\onlinecite{Ikhlas2017}. The corresponding fits of the data are represented by red lines. We extracted values of $\mu_{0} = 45~$meV, $C_\alpha = 0.004~\mathrm{V}\cdot\left(\mathrm{K~\Omega~m}\right)^{-1}$ for Mn$_{3.06}$Sn$_{0.94}$ and $\mu_{0} = 108~$meV, $C_\alpha = 0.010~\mathrm{V}\cdot\left(\mathrm{K~\Omega~m}\right)^{-1}$ for Mn$_{3.09}$Sn$_{0.91}$.}
\label{fig:Mn3Sn_Fit}
\end{figure}

Our findings show clearly that the anomalous Nernst effect, beyond the mere statement that the integrated Berry curvature near the Fermi level is finite for a given material, can also be used as a sensitive probe for the experimentally relevant strength of the Berry curvature. In this regard it is interesting to compare the obtained quantities with other Weyl materials exhibiting a similar density of states $\rho_0$. For this purpose, we performed an analogous analysis of existing transport data\cite{Ikhlas2017} of the isostructural Weyl compound \sn, see Fig.~\ref{fig:Mn3Sn_Fit}. At $T = 300$~K, $S_{xz}$ and $S_{yz}$ exhibit exactly the same step-like behavior than our data. Interestingly, $\alpha_{xz}$ in \sn is one order of magnitude smaller than in \mn. Thus, the Berry curvature in \sn at the Fermi level is significantly smaller than in \mn, since both materials possess a similar density of states $\rho_0$\cite{Yang2017}. This difference is consistent with theoretical results in Ref.~\onlinecite{Xu2018}. Furthermore, the peak of $\alpha_{xz}$, compared to \mn, is shifted to much higher temperatures. This implies, according to our considerations above, a significantly higher Weyl point energy in \sn.

The results of our analysis are shown in Fig.~\ref{fig:Mn3Sn_Fit}, where we have applied Eq.~\eqref{Fitting_formula} to fit the temperature dependence of $\alpha_{xz}$ in \sn. The parameter $C_\alpha = 0.004 \dots 0.010~\mathrm{V}\cdot\left(\mathrm{K~\Omega~m}\right)^{-1}$, which is nearly an order of magnitude lower than the corresponding value in \mn. Furthermore we obtain an energy of the lowest-lying Weyl points in the range of $\mu_0 \approx 40 \dots 100$~meV, which corresponds well to the value of $\mu = 86$~meV for \sn given by band structure calculations in Ref.~\onlinecite{Yang2017}. These findings correspond to our qualitative comparison of the magnitude and position of the maximum of $\alpha_{xz}(T)$.

The above comparison between \mn and \sn demonstrates the universal applicability of our analysis to topological materials, and thus allows to use the ANE for quantitative determination of Weyl point properties.

\section{Summary}
\label{sec:summary}

In summary, we measured the anomalous Nernst effect in \mn and developed a theoretical model to obtain quantitative information on the Weyl nodes in this material. Our analysis reveals an access to fundamental properties of Weyl systems through anomalous transverse transport. On the one hand, the anomalous Nernst effect can be used to determine the Weyl point energy as well as the momentum separation of the lowest lying Weyl points of the system. On the other hand, and most importantly, our analysis yields a measure of the Berry curvature strength at the Fermi level which is, to the best of our knowledge, not accessible through other experimental probes. In this way, our study promotes the anomalous Nernst effect as an exceptional bulk probe to detect and study Weyl physics in solid state materials.\\





\begin{acknowledgments}
This project has been supported by the Deutsche Forschungsgemeinschaft through project C07 of SFB 1143 (project-id 247310070) and priority program DFG-GRK1621. This project has received funding from the European Research Council (ERC) under the European Unions’ Horizon 2020 research and innovation program (grant agreement No 647276 – MARS – ERC-2014-CoG).\\
\end{acknowledgments}

\newpage
\begin{appendix}

\section{Peltier coefficient and its components}
\label{sec:peltier}

Since Hall- and Nernst effect show an anomalous behaviour, $\alpha_{xz}$ can be calculated using zero field values of all involved transport coefficients. The Peltier tensor $\bar{\alpha}$ can be written as
\begin{equation}
\bar{\alpha}=\bar{\sigma}\bar{S}
\end{equation}
with the conductivity tensor $\bar{\sigma}$ and the thermoelectric tensor $\bar{S}$, the former describing longitudinal and Hall conductivities and the latter Seebeck- and Nernst coefficients. Focussing on the $xz$-plane, the conductivity tensor can be expressed as the inverse of the resistivity tensor 
\begin{equation}
\bar{\sigma} = \bar{\rho}^{-1} = \frac{1}{\det\bar{\rho}}
\left(
\begin{array}{rr}
\rho_{zz} & -\rho_{xz} \\
-\rho_{zx} & \rho_{xx}
\end{array}
\right),
\end{equation}
leading to the following form of $\bar{\alpha}$:
\begin{eqnarray}
\bar{\alpha} &=& 
\left(
\begin{array}{rr}
\alpha_{xx} & \alpha_{xz} \\
\alpha_{zx} & \alpha_{zz}
\end{array} 
\right) =
\frac{1}{\rho_{xx}\rho_{zz} - \rho_{xz}\rho_{zx}} \times \nonumber \\
&&
\left(
\begin{array}{rr}
\rho_{zz}S_{xx} - \rho_{xz}S_{zx} & \rho_{zz}S_{xz} - \rho_{xz}S_{zz} \\
\rho_{xx}S_{zx} - \rho_{zx}S_{xx} & \rho_{xx}S_{zz} - \rho_{zx}S_{xz} 
\end{array}
\right).
\end{eqnarray}
With this, one can easily express $\alpha_{xz}$ as
\begin{equation}
\alpha_{xz} = \frac{\rho_{zz}S_{xz} - \rho_{xz}S_{zz}}{\rho_{xx}\rho_{zz} - \rho_{xz}\rho_{zx}},
\end{equation}
or, using the relation $\rho_{xz} = -\rho_{zx}$, as
\begin{equation}
\label{alpha_contr}
\alpha_{xz} = \frac{\rho_{zz}S_{xz} - \rho_{xz}S_{zz}}{\rho_{xx}\rho_{zz} + \rho_{xz}^{2}}\;.
\end{equation}

\begin{figure}
\includegraphics[width=1\columnwidth]{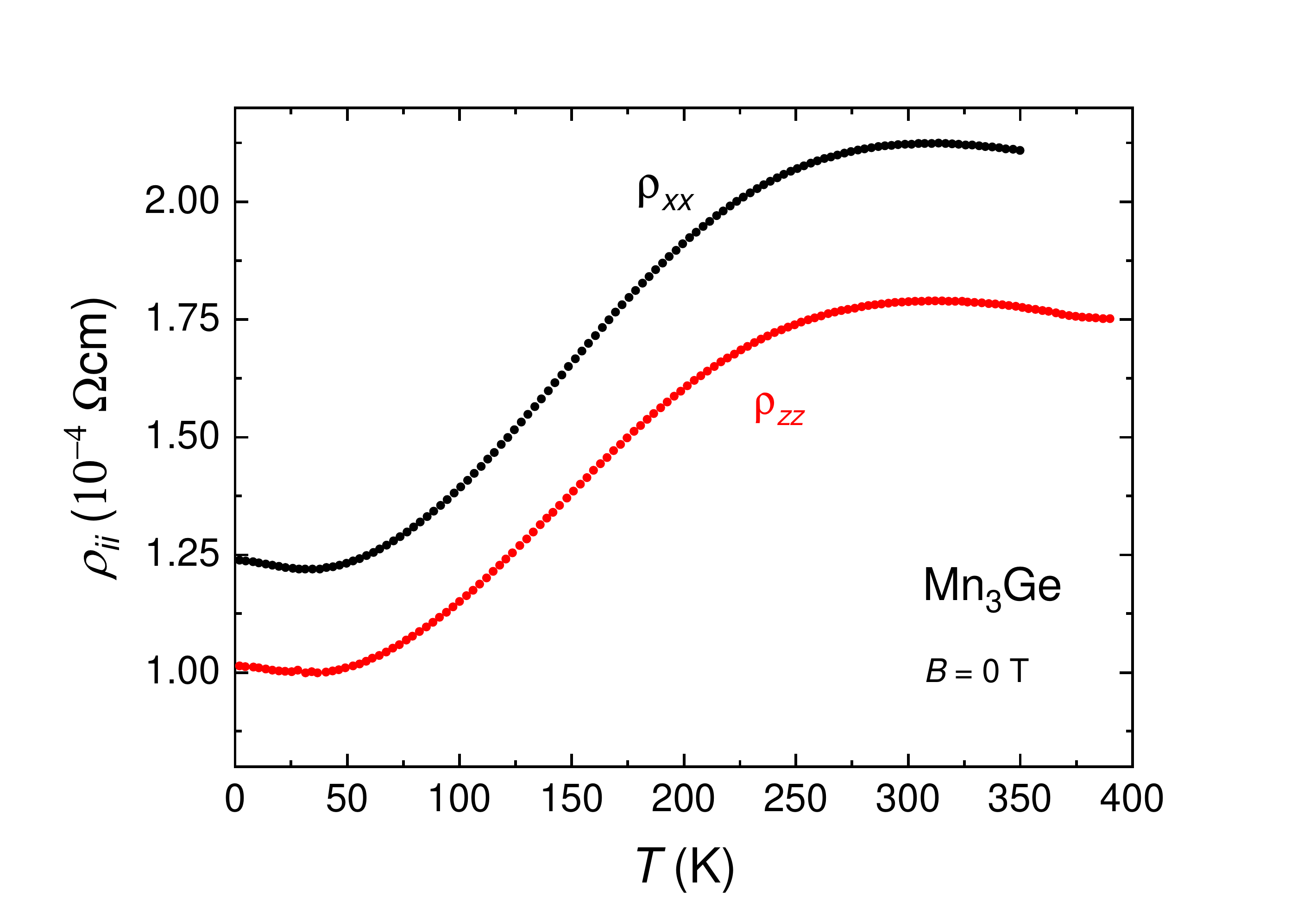}
\caption{Temperature dependent resistivity along the $x$- and $z$-direction.}
\label{fig:Supplement_Rho_ii}
\end{figure}

Electrical transport measurements were done in different configurations to calculate the Peltier coefficient $\alpha_{xz}$. The temperature dependent resistivity for in-plane and out-of-plane configurations is shown in Fig.~\ref{fig:Supplement_Rho_ii}. Fig.~\ref{fig:Sigma_Fit} shows the $xz$-component of the Hall conductivity. Both quantities are in good agreement with previous measurements\cite{Nayak2016}.

\begin{figure}
\includegraphics[width=1\columnwidth]{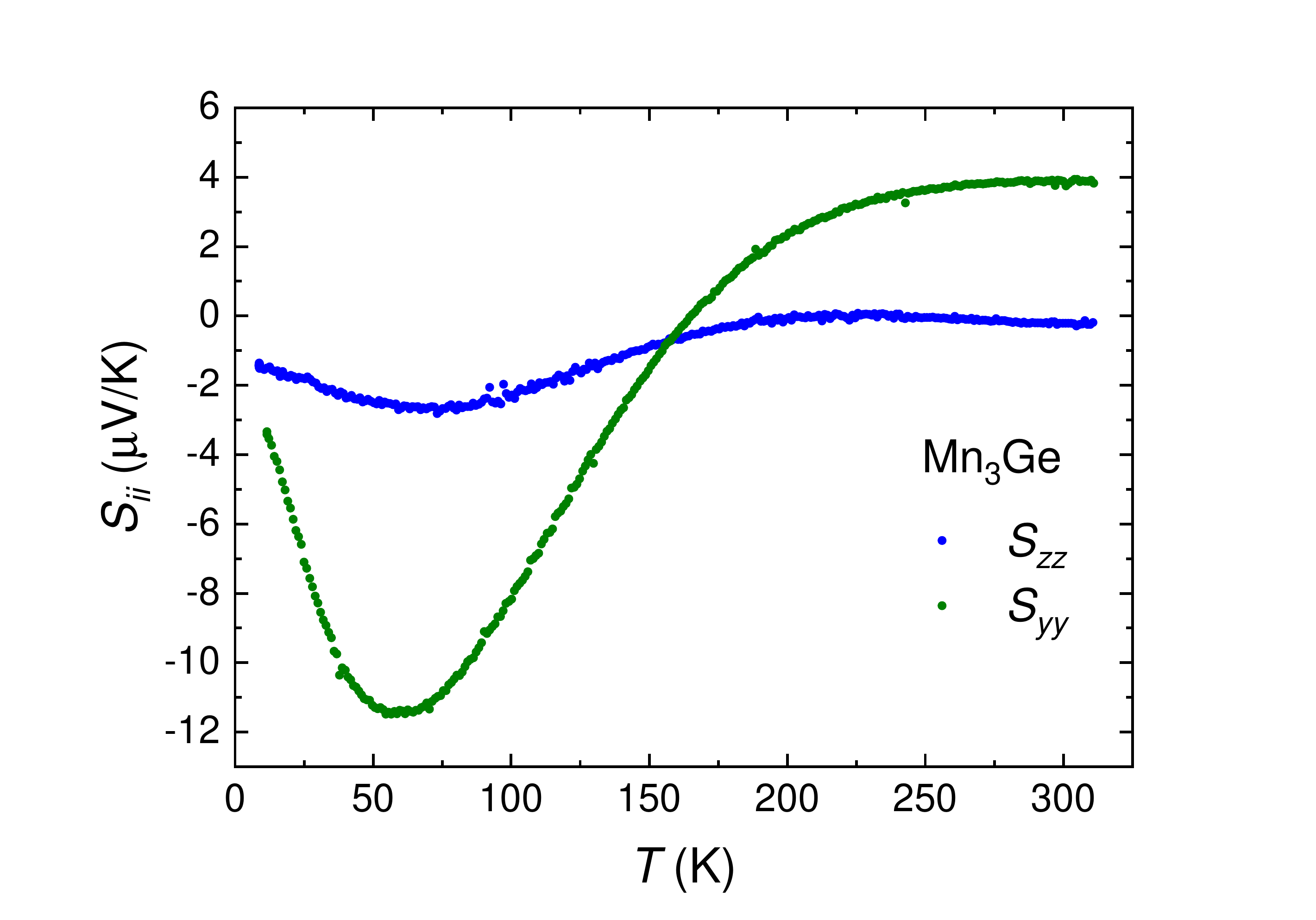}
\caption{Temperature dependent Seebeck coefficient measurement along the $y$- and $z$-direction.}
\label{fig:Supplement_S_ii}
\end{figure}

The Seebeck effect was measured using the Nernst setup with one additional electrical contact. The Seebeck coefficient $S_{ii}$ along the $y$- and $z$-axis has been studied. The data is shown in Fig.~\ref{fig:Supplement_S_ii}.

\section{Fitting of Hall effect data}
\label{sec:Hall_fit}

\begin{figure}
\includegraphics[width=1\columnwidth]{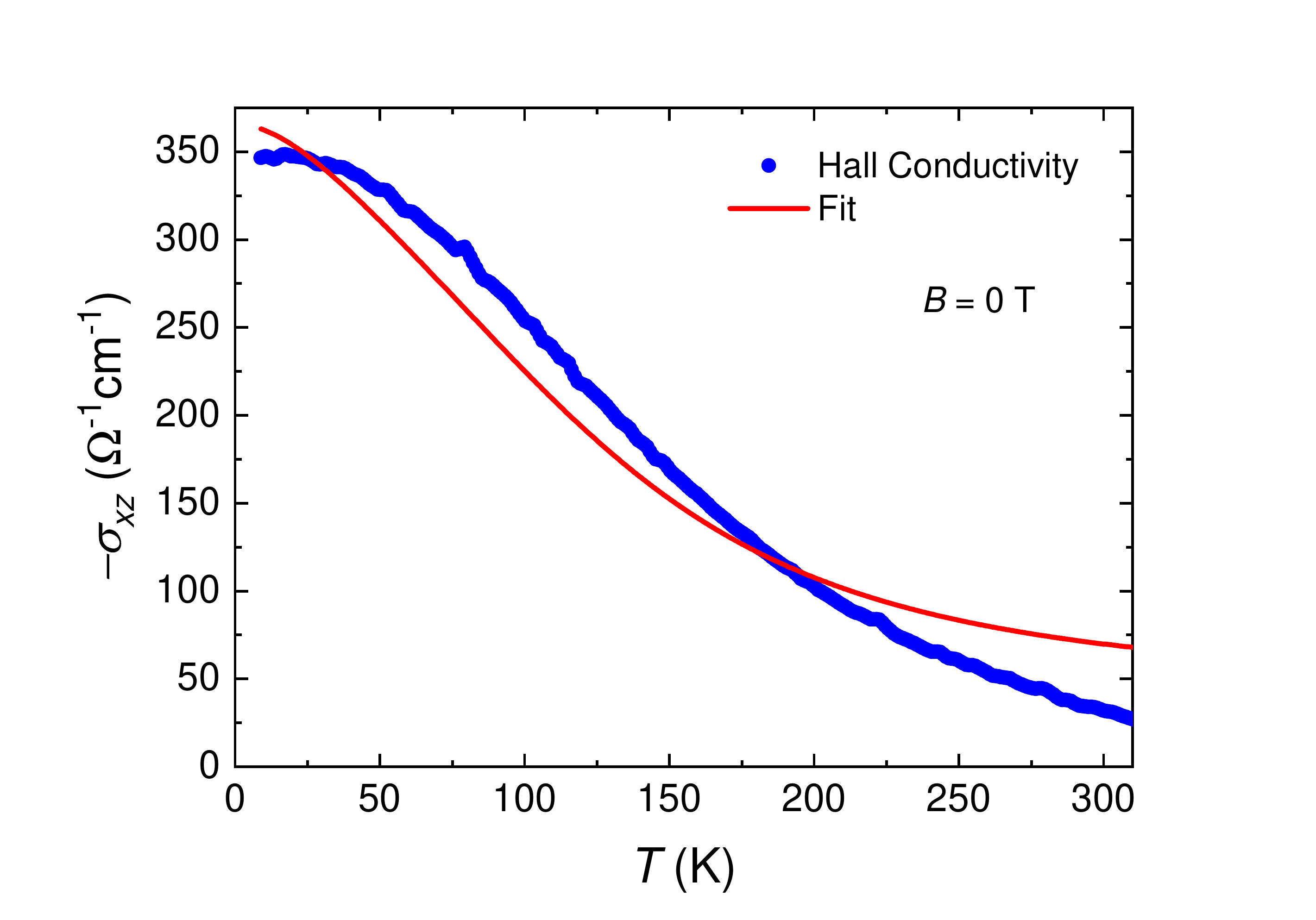}
\caption{Plot of the Hall conductivity $\sigma_{xz}$ vs. temperature. Blue dots represent experimental data. The data was fitted (red line) by using our theoretical approach as well as the fixed parameters ($\mu_0 = 8.3$~meV, $E_s = 143$~meV, and $N_W = 22.3$) extracted from the fit of $\alpha_{xz}$.}
\label{fig:Sigma_Fit}
\end{figure}

Using the same arguments as described in Sec.~\ref{sec:theory} we have derived a similar fitting formula for the anomalous Hall coefficient $\sigma_{xz}$, based on the usual expression 
\begin{equation}
\sigma_{xz}=\frac{e^2}{\hbar} \int \frac{d^3k}{(2\pi)^3} \Omega^{y}({\bf k}) F(\varepsilon_{\bf k})
\label{sigma_xy}
\end{equation}
as obtained from the Boltzmann transport theory.
The evaluation of the momentum integration results in the following formula,  
\begin{eqnarray}
\sigma_{xz} &=& C_\sigma k_B T \left[ 2 \ln \left(1 + e^{(E_s/2 - \mu)/k_BT} \right) \right.  \nonumber \\
&&- \left. \ln \left(1 + e^{(- E_s/2 - \mu)/k_BT} \right) \right],
\label{sigma_xy_fitting}
\end{eqnarray}
where $C_{\sigma}$ is a specific fitting parameter of $\sigma_{xz}$. This formula can be used, together with the parameters obtained by fitting the temperature dependence of $\alpha_{xz}$ ($\mu_0 = 6.6 \pm 0.7$~meV, $E_S = 90 \pm 25$~meV, and $N_W = 17.8 \pm 2.2$), to fit the Hall conductivity $\sigma_{xz}$. The parameters $\mu_0$, $E_S$ and $N_W$ were allowed to vary inside the corresponding errorbar. The result is displayed in Fig.~\ref{fig:Sigma_Fit}, the remaining parameter $C_{\sigma} = -3.6 \pm 0.3~\left(\mathrm{\Omega~cm~meV}\right)^{-1}$. As one can see the fit works less well than for the Peltier coefficient in Fig.~\ref{fig:S_ij(T)}(b) for the following reason. Due to the presence of the Fermi function the momentum integration in Eq.~\eqref{sigma_xy_fitting} is not restricted to states close to the Fermi level. Therefore, an additional approximation is needed to be included to perform the momentum integration analytically. More specifically, due to the entropy density in the momentum integral of the anomalous Peltier coefficient (at low temperature) it is solely determined by states close to the Fermi level. This is however not the case for the anomalous Hall conductivity where also states deep in the occupied region of the Fermi sea contribute. We emphasize at this point that such additional approximation rather affects the temperature dependence than the overall magnitude. Therefore we believe that the coefficients $C_\alpha$ and $C_\sigma$, which are the responsible parameters of the overall magnitude, should be rather unaffected by the discussed approximation. Hence, since these parameters are the only ones which explicitly contain the Berry curvature, we argue that our main conclusion regarding the strength of the Berry curvature is still valid despite the fact that the temperature fit of the Hall coefficient is not as good as that for the Peltier coefficient.

\section{Low-field Nernst signal and magnetization behavior}
\label{sec:magn}

\begin{figure}
\includegraphics[width=0.9\columnwidth]{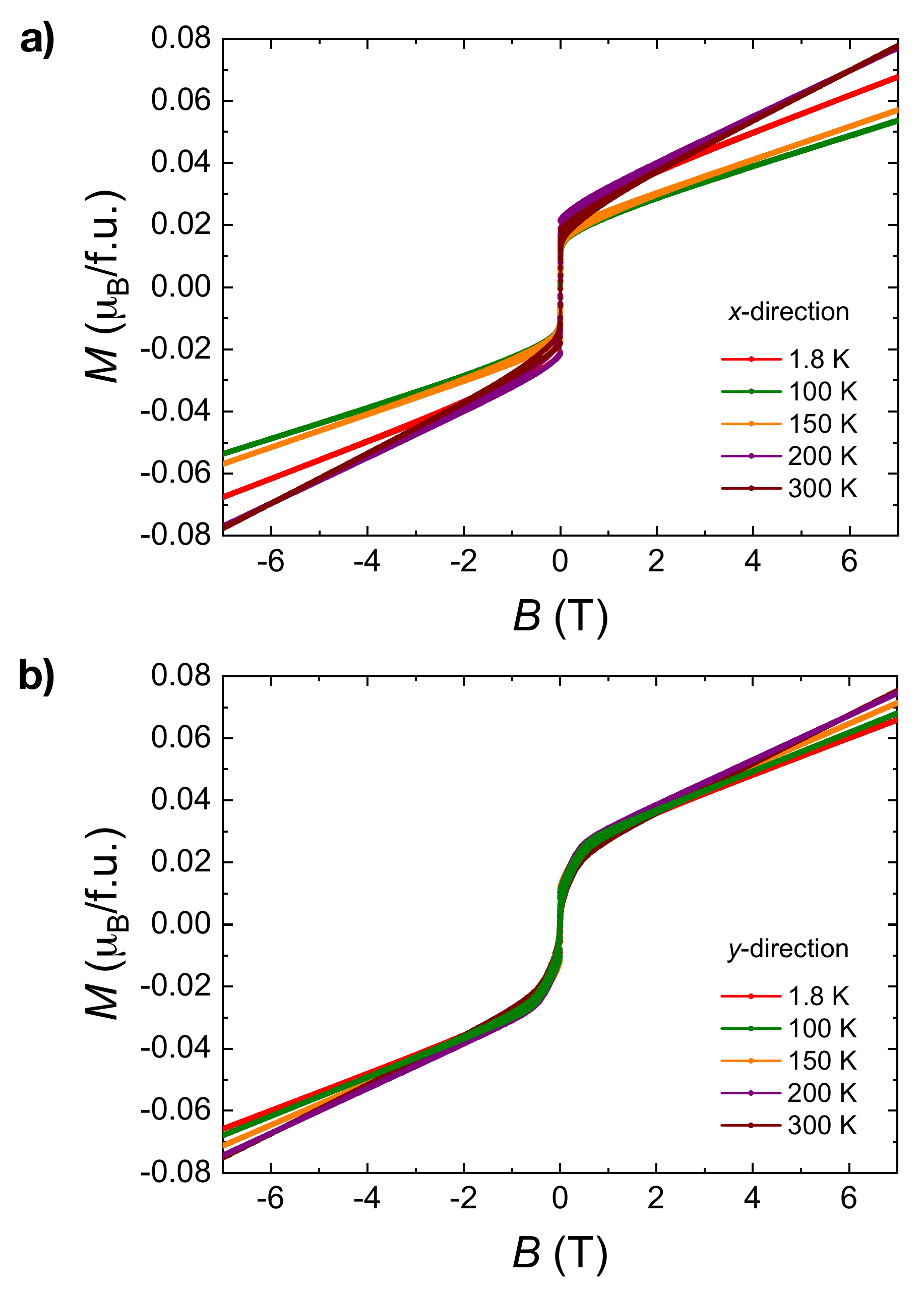}
\caption{Magnetic moment of \mn in \textbf{a)} $x$-direction and \textbf{b)} $y$-direction. The magnetization has been corrected for geometry effects. However, demagnetization corrections are small and have not been applied. $B$ denotes the external magnetic field.}
\label{fig:Appendix_Magn(B)}
\end{figure}

\begin{figure*}
\includegraphics[width=1\textwidth]{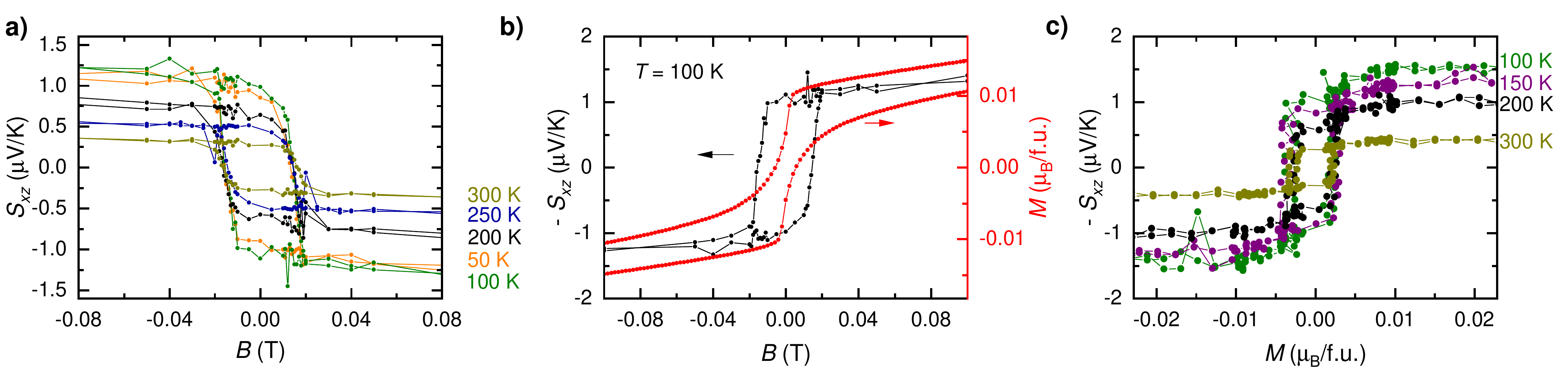}
\caption{\textbf{a)} Zoom-in of $S_{xz}\left(B\right)$ with a clearly visible hysteresis of the Nernst signal at different selected temperatures. \textbf{b)} Comparison of the hysteresis curves of $S_{xz}\left(B\right)$ and the magnetization $M$ in y-direction at $T = 100~$K. $B$ denotes the external magnetic field. \textbf{c)} $S_{xz}$ vs. $M$. There is no obvious scaling of the anomalous Nernst signal on the magnetization of the sample.}
\label{fig:S_ij(M)}
\end{figure*}

In addition to the discussed findings of the main text we would like to mention the low-field region of the Nernst data. As highlighted in Fig.~\ref{fig:S_ij(M)}, both the $S_{xz}$ vs $B$ and the $S_{yz}$ vs $B$ curves (the latter are not shown) exhibit a hysteresis cycle which remains almost unaltered from 5~K up to room temperature. This cycle is rectangular-shaped and closes at around 20~mT, in agreement with the previous report on \sn\cite{Ikhlas2017}. Remarkably, the cycle exhibits a total extension of around $\Delta S_{xz} = 2~\mu$V/K at 100~K (even overcoming the peak value of $\Delta S_{xz} = 0.7~\mu$V/K observed in \sn\cite{Ikhlas2017}) with negligible net magnetization. This exceptionally large value underpins that non-collinear antiferromagnets such as \mn and \sn may constitute a new material paradigm in the field of spintronic\cite{Kimata2019, Nakatsuji2015, Nayak2015, Park2011, Marti2014} and thermoelectric technologies\cite{Ikhlas2017, Mizukami2013}. 

It is interesting to compare this low-field behavior of the ANE with the DC magnetization. The latter was measured as a function of the temperature and magnetic field by means of a quantum interference device magnetometer (SQUID-VSM) by Quantum Design. For the purpose of a direct comparison with the Nernst effect measurements, the magnetization curves where obtained by applying an external magnetic field in the direction(s) [2-1-10] (and [01-10]). In order to probe the field dependent magnetization, M(B) (Fig. \ref{fig:Appendix_Magn(B)}) was measured upon sweeping the magnetic field between -7~T to 7~T at constant temperatures 1.8, 50, 100, 150, 200, 250, 300~K. Given the large magnitude of the measured magnetic signal, the small background correction due to the diamagnetic contribution of the glue GE-Varnish used to fix the sample has been neglected.

It is well-known that a magnetization (even for a relatively small value) may enhance the ANE as it is the case in ferromagnets where the anomalous Hall conductivity is usually assumed to be proportional to the magnetization of the magnetic material. Interestingly, such a proportionality between the ANE and the magnetization $M$ is absent in \mn, see Fig.~\ref{fig:S_ij(M)}(b). Even if $M$ somehow reproduces the overall hysteresis shape in the low field region it closes the cycle at much higher fields with respect to $S_{xz}$. Furthermore, while $S_{xz}$ is anomalous and as such stays constant with increasing $B$, $M$ undergoes an almost linear drift. We also verified the absence of a scaling of the two quantities by plotting $S_{xz}$ vs.~$M$ in Fig.~\ref{fig:S_ij(M)}(c).

\end{appendix}


\begin{thebibliography}{38}%
\makeatletter
\providecommand \@ifxundefined [1]{%
 \@ifx{#1\undefined}
}%
\providecommand \@ifnum [1]{%
 \ifnum #1\expandafter \@firstoftwo
 \else \expandafter \@secondoftwo
 \fi
}%
\providecommand \@ifx [1]{%
 \ifx #1\expandafter \@firstoftwo
 \else \expandafter \@secondoftwo
 \fi
}%
\providecommand \natexlab [1]{#1}%
\providecommand \enquote  [1]{``#1''}%
\providecommand \bibnamefont  [1]{#1}%
\providecommand \bibfnamefont [1]{#1}%
\providecommand \citenamefont [1]{#1}%
\providecommand \href@noop [0]{\@secondoftwo}%
\providecommand \href [0]{\begingroup \@sanitize@url \@href}%
\providecommand \@href[1]{\@@startlink{#1}\@@href}%
\providecommand \@@href[1]{\endgroup#1\@@endlink}%
\providecommand \@sanitize@url [0]{\catcode `\\12\catcode `\$12\catcode
  `\&12\catcode `\#12\catcode `\^12\catcode `\_12\catcode `\%12\relax}%
\providecommand \@@startlink[1]{}%
\providecommand \@@endlink[0]{}%
\providecommand \url  [0]{\begingroup\@sanitize@url \@url }%
\providecommand \@url [1]{\endgroup\@href {#1}{\urlprefix }}%
\providecommand \urlprefix  [0]{URL }%
\providecommand \Eprint [0]{\href }%
\providecommand \doibase [0]{http://dx.doi.org/}%
\providecommand \selectlanguage [0]{\@gobble}%
\providecommand \bibinfo  [0]{\@secondoftwo}%
\providecommand \bibfield  [0]{\@secondoftwo}%
\providecommand \translation [1]{[#1]}%
\providecommand \BibitemOpen [0]{}%
\providecommand \bibitemStop [0]{}%
\providecommand \bibitemNoStop [0]{.\EOS\space}%
\providecommand \EOS [0]{\spacefactor3000\relax}%
\providecommand \BibitemShut  [1]{\csname bibitem#1\endcsname}%
\let\auto@bib@innerbib\@empty
\bibitem [{\citenamefont {Wan}\ \emph {et~al.}(2011)\citenamefont {Wan},
  \citenamefont {Turner}, \citenamefont {Vishwanath},\ and\ \citenamefont
  {Savrasov}}]{Wan2011}%
  \BibitemOpen
  \bibfield  {author} {\bibinfo {author} {\bibfnamefont {X.}~\bibnamefont
  {Wan}}, \bibinfo {author} {\bibfnamefont {A.~M.}\ \bibnamefont {Turner}},
  \bibinfo {author} {\bibfnamefont {A.}~\bibnamefont {Vishwanath}}, \ and\
  \bibinfo {author} {\bibfnamefont {S.~Y.}\ \bibnamefont {Savrasov}},\ }\href
  {\doibase 10.1103/PhysRevB.83.205101} {\bibfield  {journal} {\bibinfo
  {journal} {Phys. Rev. B}\ }\textbf {\bibinfo {volume} {83}},\ \bibinfo
  {pages} {205101} (\bibinfo {year} {2011})}\BibitemShut {NoStop}%
\bibitem [{\citenamefont {Yang}\ \emph {et~al.}(2015)\citenamefont {Yang},
  \citenamefont {Liu}, \citenamefont {Sun}, \citenamefont {Peng}, \citenamefont
  {Yang}, \citenamefont {Zhang}, \citenamefont {Zhou}, \citenamefont {Zhang},
  \citenamefont {Guo}, \citenamefont {Rahn}, \citenamefont {Prabhakaran},
  \citenamefont {Hussain}, \citenamefont {Mo}, \citenamefont {Felser},
  \citenamefont {Yan},\ and\ \citenamefont {Chen}}]{Yang2015}%
  \BibitemOpen
  \bibfield  {author} {\bibinfo {author} {\bibfnamefont {L.~X.}\ \bibnamefont
  {Yang}}, \bibinfo {author} {\bibfnamefont {Z.~K.}\ \bibnamefont {Liu}},
  \bibinfo {author} {\bibfnamefont {Y.}~\bibnamefont {Sun}}, \bibinfo {author}
  {\bibfnamefont {H.}~\bibnamefont {Peng}}, \bibinfo {author} {\bibfnamefont
  {H.~F.}\ \bibnamefont {Yang}}, \bibinfo {author} {\bibfnamefont
  {T.}~\bibnamefont {Zhang}}, \bibinfo {author} {\bibfnamefont
  {B.}~\bibnamefont {Zhou}}, \bibinfo {author} {\bibfnamefont {Y.}~\bibnamefont
  {Zhang}}, \bibinfo {author} {\bibfnamefont {Y.~F.}\ \bibnamefont {Guo}},
  \bibinfo {author} {\bibfnamefont {M.}~\bibnamefont {Rahn}}, \bibinfo {author}
  {\bibfnamefont {D.}~\bibnamefont {Prabhakaran}}, \bibinfo {author}
  {\bibfnamefont {Z.}~\bibnamefont {Hussain}}, \bibinfo {author} {\bibfnamefont
  {S.-K.}\ \bibnamefont {Mo}}, \bibinfo {author} {\bibfnamefont
  {C.}~\bibnamefont {Felser}}, \bibinfo {author} {\bibfnamefont
  {B.}~\bibnamefont {Yan}}, \ and\ \bibinfo {author} {\bibfnamefont {Y.~L.}\
  \bibnamefont {Chen}},\ }\href {https://doi.org/10.1038/nphys3425} {\bibfield
  {journal} {\bibinfo  {journal} {Nature Physics}\ }\textbf {\bibinfo {volume}
  {11}},\ \bibinfo {pages} {728 EP } (\bibinfo {year} {2015})}\BibitemShut
  {NoStop}%
\bibitem [{\citenamefont {Nakatsuji}\ \emph {et~al.}(2015)\citenamefont
  {Nakatsuji}, \citenamefont {Kiyohara},\ and\ \citenamefont
  {Higo}}]{Nakatsuji2015}%
  \BibitemOpen
  \bibfield  {author} {\bibinfo {author} {\bibfnamefont {S.}~\bibnamefont
  {Nakatsuji}}, \bibinfo {author} {\bibfnamefont {N.}~\bibnamefont {Kiyohara}},
  \ and\ \bibinfo {author} {\bibfnamefont {T.}~\bibnamefont {Higo}},\ }\href
  {http://dx.doi.org/10.1038/nature15723} {\bibfield  {journal} {\bibinfo
  {journal} {Nature}\ }\textbf {\bibinfo {volume} {527}},\ \bibinfo {pages}
  {212 EP } (\bibinfo {year} {2015})}\BibitemShut {NoStop}%
\bibitem [{\citenamefont {Armitage}\ \emph {et~al.}(2018)\citenamefont
  {Armitage}, \citenamefont {Mele},\ and\ \citenamefont
  {Vishwanath}}]{Armitage2018}%
  \BibitemOpen
  \bibfield  {author} {\bibinfo {author} {\bibfnamefont {N.~P.}\ \bibnamefont
  {Armitage}}, \bibinfo {author} {\bibfnamefont {E.~J.}\ \bibnamefont {Mele}},
  \ and\ \bibinfo {author} {\bibfnamefont {A.}~\bibnamefont {Vishwanath}},\
  }\href {\doibase 10.1103/RevModPhys.90.015001} {\bibfield  {journal}
  {\bibinfo  {journal} {Rev. Mod. Phys.}\ }\textbf {\bibinfo {volume} {90}},\
  \bibinfo {pages} {015001} (\bibinfo {year} {2018})}\BibitemShut {NoStop}%
\bibitem [{\citenamefont {Kimata}\ \emph {et~al.}(2019)\citenamefont {Kimata},
  \citenamefont {Chen}, \citenamefont {Kondou}, \citenamefont {Sugimoto},
  \citenamefont {Muduli}, \citenamefont {Ikhlas}, \citenamefont {Omori},
  \citenamefont {Tomita}, \citenamefont {MacDonald}, \citenamefont
  {Nakatsuji},\ and\ \citenamefont {Otani}}]{Kimata2019}%
  \BibitemOpen
  \bibfield  {author} {\bibinfo {author} {\bibfnamefont {M.}~\bibnamefont
  {Kimata}}, \bibinfo {author} {\bibfnamefont {H.}~\bibnamefont {Chen}},
  \bibinfo {author} {\bibfnamefont {K.}~\bibnamefont {Kondou}}, \bibinfo
  {author} {\bibfnamefont {S.}~\bibnamefont {Sugimoto}}, \bibinfo {author}
  {\bibfnamefont {P.~K.}\ \bibnamefont {Muduli}}, \bibinfo {author}
  {\bibfnamefont {M.}~\bibnamefont {Ikhlas}}, \bibinfo {author} {\bibfnamefont
  {Y.}~\bibnamefont {Omori}}, \bibinfo {author} {\bibfnamefont
  {T.}~\bibnamefont {Tomita}}, \bibinfo {author} {\bibfnamefont {A.~H.}\
  \bibnamefont {MacDonald}}, \bibinfo {author} {\bibfnamefont {S.}~\bibnamefont
  {Nakatsuji}}, \ and\ \bibinfo {author} {\bibfnamefont {Y.}~\bibnamefont
  {Otani}},\ }\href {\doibase 10.1038/s41586-018-0853-0} {\bibfield  {journal}
  {\bibinfo  {journal} {Nature}\ } (\bibinfo {year} {2019}),\
  10.1038/s41586-018-0853-0}\BibitemShut {NoStop}%
\bibitem [{\citenamefont {Kurebayashi}\ \emph {et~al.}(2014)\citenamefont
  {Kurebayashi}, \citenamefont {Sinova}, \citenamefont {Fang}, \citenamefont
  {Irvine}, \citenamefont {Skinner}, \citenamefont {Wunderlich}, \citenamefont
  {Nov{\'a}k}, \citenamefont {Campion}, \citenamefont {Gallagher},
  \citenamefont {Vehstedt}, \citenamefont {Z{\^a}rbo}, \citenamefont
  {V{\'y}born{\'y}}, \citenamefont {Ferguson},\ and\ \citenamefont
  {Jungwirth}}]{Kurebayashi2014}%
  \BibitemOpen
  \bibfield  {author} {\bibinfo {author} {\bibfnamefont {H.}~\bibnamefont
  {Kurebayashi}}, \bibinfo {author} {\bibfnamefont {J.}~\bibnamefont {Sinova}},
  \bibinfo {author} {\bibfnamefont {D.}~\bibnamefont {Fang}}, \bibinfo {author}
  {\bibfnamefont {A.~C.}\ \bibnamefont {Irvine}}, \bibinfo {author}
  {\bibfnamefont {T.~D.}\ \bibnamefont {Skinner}}, \bibinfo {author}
  {\bibfnamefont {J.}~\bibnamefont {Wunderlich}}, \bibinfo {author}
  {\bibfnamefont {V.}~\bibnamefont {Nov{\'a}k}}, \bibinfo {author}
  {\bibfnamefont {R.~P.}\ \bibnamefont {Campion}}, \bibinfo {author}
  {\bibfnamefont {B.~L.}\ \bibnamefont {Gallagher}}, \bibinfo {author}
  {\bibfnamefont {E.~K.}\ \bibnamefont {Vehstedt}}, \bibinfo {author}
  {\bibfnamefont {L.~P.}\ \bibnamefont {Z{\^a}rbo}}, \bibinfo {author}
  {\bibfnamefont {K.}~\bibnamefont {V{\'y}born{\'y}}}, \bibinfo {author}
  {\bibfnamefont {A.~J.}\ \bibnamefont {Ferguson}}, \ and\ \bibinfo {author}
  {\bibfnamefont {T.}~\bibnamefont {Jungwirth}},\ }\href
  {https://doi.org/10.1038/nnano.2014.15} {\bibfield  {journal} {\bibinfo
  {journal} {Nature Nanotechnology}\ }\textbf {\bibinfo {volume} {9}},\
  \bibinfo {pages} {211 EP } (\bibinfo {year} {2014})},\ \bibinfo {note}
  {article}\BibitemShut {NoStop}%
\bibitem [{\citenamefont {Yale}\ \emph {et~al.}(2016)\citenamefont {Yale},
  \citenamefont {Heremans}, \citenamefont {Zhou}, \citenamefont {Auer},
  \citenamefont {Burkard},\ and\ \citenamefont {Awschalom}}]{Yale2016}%
  \BibitemOpen
  \bibfield  {author} {\bibinfo {author} {\bibfnamefont {C.~G.}\ \bibnamefont
  {Yale}}, \bibinfo {author} {\bibfnamefont {F.~J.}\ \bibnamefont {Heremans}},
  \bibinfo {author} {\bibfnamefont {B.~B.}\ \bibnamefont {Zhou}}, \bibinfo
  {author} {\bibfnamefont {A.}~\bibnamefont {Auer}}, \bibinfo {author}
  {\bibfnamefont {G.}~\bibnamefont {Burkard}}, \ and\ \bibinfo {author}
  {\bibfnamefont {D.~D.}\ \bibnamefont {Awschalom}},\ }\href
  {https://doi.org/10.1038/nphoton.2015.278} {\bibfield  {journal} {\bibinfo
  {journal} {Nature Photonics}\ }\textbf {\bibinfo {volume} {10}},\ \bibinfo
  {pages} {184 EP } (\bibinfo {year} {2016})},\ \bibinfo {note}
  {article}\BibitemShut {NoStop}%
\bibitem [{\citenamefont {Xiao}\ \emph {et~al.}(2010)\citenamefont {Xiao},
  \citenamefont {Chang},\ and\ \citenamefont {Niu}}]{Xiao2010}%
  \BibitemOpen
  \bibfield  {author} {\bibinfo {author} {\bibfnamefont {D.}~\bibnamefont
  {Xiao}}, \bibinfo {author} {\bibfnamefont {M.-C.}\ \bibnamefont {Chang}}, \
  and\ \bibinfo {author} {\bibfnamefont {Q.}~\bibnamefont {Niu}},\ }\href
  {\doibase 10.1103/RevModPhys.82.1959} {\bibfield  {journal} {\bibinfo
  {journal} {Rev. Mod. Phys.}\ }\textbf {\bibinfo {volume} {82}},\ \bibinfo
  {pages} {1959} (\bibinfo {year} {2010})}\BibitemShut {NoStop}%
\bibitem [{\citenamefont {Sundaram}\ and\ \citenamefont
  {Niu}(1999)}]{Sundaram1999}%
  \BibitemOpen
  \bibfield  {author} {\bibinfo {author} {\bibfnamefont {G.}~\bibnamefont
  {Sundaram}}\ and\ \bibinfo {author} {\bibfnamefont {Q.}~\bibnamefont {Niu}},\
  }\href {\doibase 10.1103/PhysRevB.59.14915} {\bibfield  {journal} {\bibinfo
  {journal} {Phys. Rev. B}\ }\textbf {\bibinfo {volume} {59}},\ \bibinfo
  {pages} {14915} (\bibinfo {year} {1999})}\BibitemShut {NoStop}%
\bibitem [{\citenamefont {Nagaosa}\ \emph {et~al.}(2010)\citenamefont
  {Nagaosa}, \citenamefont {Sinova}, \citenamefont {Onoda}, \citenamefont
  {MacDonald},\ and\ \citenamefont {Ong}}]{Nagaosa2010}%
  \BibitemOpen
  \bibfield  {author} {\bibinfo {author} {\bibfnamefont {N.}~\bibnamefont
  {Nagaosa}}, \bibinfo {author} {\bibfnamefont {J.}~\bibnamefont {Sinova}},
  \bibinfo {author} {\bibfnamefont {S.}~\bibnamefont {Onoda}}, \bibinfo
  {author} {\bibfnamefont {A.~H.}\ \bibnamefont {MacDonald}}, \ and\ \bibinfo
  {author} {\bibfnamefont {N.~P.}\ \bibnamefont {Ong}},\ }\href {\doibase
  10.1103/RevModPhys.82.1539} {\bibfield  {journal} {\bibinfo  {journal} {Rev.
  Mod. Phys.}\ }\textbf {\bibinfo {volume} {82}},\ \bibinfo {pages} {1539}
  (\bibinfo {year} {2010})}\BibitemShut {NoStop}%
\bibitem [{\citenamefont {Noky}\ \emph {et~al.}(2018)\citenamefont {Noky},
  \citenamefont {Gayles}, \citenamefont {Felser},\ and\ \citenamefont
  {Sun}}]{Noky2018}%
  \BibitemOpen
  \bibfield  {author} {\bibinfo {author} {\bibfnamefont {J.}~\bibnamefont
  {Noky}}, \bibinfo {author} {\bibfnamefont {J.}~\bibnamefont {Gayles}},
  \bibinfo {author} {\bibfnamefont {C.}~\bibnamefont {Felser}}, \ and\ \bibinfo
  {author} {\bibfnamefont {Y.}~\bibnamefont {Sun}},\ }\href {\doibase
  10.1103/PhysRevB.97.220405} {\bibfield  {journal} {\bibinfo  {journal} {Phys.
  Rev. B}\ }\textbf {\bibinfo {volume} {97}},\ \bibinfo {pages} {220405(R)}
  (\bibinfo {year} {2018})}\BibitemShut {NoStop}%
\bibitem [{\citenamefont {Caglieris}\ \emph {et~al.}(2018)\citenamefont
  {Caglieris}, \citenamefont {Wuttke}, \citenamefont {Sykora}, \citenamefont
  {S\"uss}, \citenamefont {Shekhar}, \citenamefont {Felser}, \citenamefont
  {B\"uchner},\ and\ \citenamefont {Hess}}]{Caglieris2018}%
  \BibitemOpen
  \bibfield  {author} {\bibinfo {author} {\bibfnamefont {F.}~\bibnamefont
  {Caglieris}}, \bibinfo {author} {\bibfnamefont {C.}~\bibnamefont {Wuttke}},
  \bibinfo {author} {\bibfnamefont {S.}~\bibnamefont {Sykora}}, \bibinfo
  {author} {\bibfnamefont {V.}~\bibnamefont {S\"uss}}, \bibinfo {author}
  {\bibfnamefont {C.}~\bibnamefont {Shekhar}}, \bibinfo {author} {\bibfnamefont
  {C.}~\bibnamefont {Felser}}, \bibinfo {author} {\bibfnamefont
  {B.}~\bibnamefont {B\"uchner}}, \ and\ \bibinfo {author} {\bibfnamefont
  {C.}~\bibnamefont {Hess}},\ }\href {\doibase 10.1103/PhysRevB.98.201107}
  {\bibfield  {journal} {\bibinfo  {journal} {Phys. Rev. B}\ }\textbf {\bibinfo
  {volume} {98}},\ \bibinfo {pages} {201107(R)} (\bibinfo {year}
  {2018})}\BibitemShut {NoStop}%
\bibitem [{\citenamefont {Watzman}\ \emph {et~al.}(2018)\citenamefont
  {Watzman}, \citenamefont {McCormick}, \citenamefont {Shekhar}, \citenamefont
  {Wu}, \citenamefont {Sun}, \citenamefont {Prakash}, \citenamefont {Felser},
  \citenamefont {Trivedi},\ and\ \citenamefont {Heremans}}]{Watzman2018}%
  \BibitemOpen
  \bibfield  {author} {\bibinfo {author} {\bibfnamefont {S.~J.}\ \bibnamefont
  {Watzman}}, \bibinfo {author} {\bibfnamefont {T.~M.}\ \bibnamefont
  {McCormick}}, \bibinfo {author} {\bibfnamefont {C.}~\bibnamefont {Shekhar}},
  \bibinfo {author} {\bibfnamefont {S.-C.}\ \bibnamefont {Wu}}, \bibinfo
  {author} {\bibfnamefont {Y.}~\bibnamefont {Sun}}, \bibinfo {author}
  {\bibfnamefont {A.}~\bibnamefont {Prakash}}, \bibinfo {author} {\bibfnamefont
  {C.}~\bibnamefont {Felser}}, \bibinfo {author} {\bibfnamefont
  {N.}~\bibnamefont {Trivedi}}, \ and\ \bibinfo {author} {\bibfnamefont
  {J.~P.}\ \bibnamefont {Heremans}},\ }\href {\doibase
  10.1103/PhysRevB.97.161404} {\bibfield  {journal} {\bibinfo  {journal} {Phys.
  Rev. B}\ }\textbf {\bibinfo {volume} {97}},\ \bibinfo {pages} {161404(R)}
  (\bibinfo {year} {2018})}\BibitemShut {NoStop}%
\bibitem [{\citenamefont {Sakai}\ \emph {et~al.}(2018)\citenamefont {Sakai},
  \citenamefont {Mizuta}, \citenamefont {Nugroho}, \citenamefont {Sihombing},
  \citenamefont {Koretsune}, \citenamefont {Suzuki}, \citenamefont {Takemori},
  \citenamefont {Ishii}, \citenamefont {Nishio-Hamane}, \citenamefont {Arita},
  \citenamefont {Goswami},\ and\ \citenamefont {Nakatsuji}}]{Sakai2018}%
  \BibitemOpen
  \bibfield  {author} {\bibinfo {author} {\bibfnamefont {A.}~\bibnamefont
  {Sakai}}, \bibinfo {author} {\bibfnamefont {Y.~P.}\ \bibnamefont {Mizuta}},
  \bibinfo {author} {\bibfnamefont {A.~A.}\ \bibnamefont {Nugroho}}, \bibinfo
  {author} {\bibfnamefont {R.}~\bibnamefont {Sihombing}}, \bibinfo {author}
  {\bibfnamefont {T.}~\bibnamefont {Koretsune}}, \bibinfo {author}
  {\bibfnamefont {M.-T.}\ \bibnamefont {Suzuki}}, \bibinfo {author}
  {\bibfnamefont {N.}~\bibnamefont {Takemori}}, \bibinfo {author}
  {\bibfnamefont {R.}~\bibnamefont {Ishii}}, \bibinfo {author} {\bibfnamefont
  {D.}~\bibnamefont {Nishio-Hamane}}, \bibinfo {author} {\bibfnamefont
  {R.}~\bibnamefont {Arita}}, \bibinfo {author} {\bibfnamefont
  {P.}~\bibnamefont {Goswami}}, \ and\ \bibinfo {author} {\bibfnamefont
  {S.}~\bibnamefont {Nakatsuji}},\ }\href {\doibase 10.1038/s41567-018-0225-6}
  {\bibfield  {journal} {\bibinfo  {journal} {Nature Physics}\ }\textbf
  {\bibinfo {volume} {14}},\ \bibinfo {pages} {1119} (\bibinfo {year}
  {2018})}\BibitemShut {NoStop}%
\bibitem [{\citenamefont {Li}\ \emph {et~al.}(2017)\citenamefont {Li},
  \citenamefont {Xu}, \citenamefont {Ding}, \citenamefont {Wang}, \citenamefont
  {Shen}, \citenamefont {Lu}, \citenamefont {Zhu},\ and\ \citenamefont
  {Behnia}}]{Li2017}%
  \BibitemOpen
  \bibfield  {author} {\bibinfo {author} {\bibfnamefont {X.}~\bibnamefont
  {Li}}, \bibinfo {author} {\bibfnamefont {L.}~\bibnamefont {Xu}}, \bibinfo
  {author} {\bibfnamefont {L.}~\bibnamefont {Ding}}, \bibinfo {author}
  {\bibfnamefont {J.}~\bibnamefont {Wang}}, \bibinfo {author} {\bibfnamefont
  {M.}~\bibnamefont {Shen}}, \bibinfo {author} {\bibfnamefont {X.}~\bibnamefont
  {Lu}}, \bibinfo {author} {\bibfnamefont {Z.}~\bibnamefont {Zhu}}, \ and\
  \bibinfo {author} {\bibfnamefont {K.}~\bibnamefont {Behnia}},\ }\href
  {\doibase 10.1103/PhysRevLett.119.056601} {\bibfield  {journal} {\bibinfo
  {journal} {Phys. Rev. Lett.}\ }\textbf {\bibinfo {volume} {119}},\ \bibinfo
  {pages} {056601} (\bibinfo {year} {2017})}\BibitemShut {NoStop}%
\bibitem [{\citenamefont {Nayak}\ \emph {et~al.}(2016)\citenamefont {Nayak},
  \citenamefont {Fischer}, \citenamefont {Sun}, \citenamefont {Yan},
  \citenamefont {Karel}, \citenamefont {Komarek}, \citenamefont {Shekhar},
  \citenamefont {Kumar}, \citenamefont {Schnelle}, \citenamefont {K{\"u}bler},
  \citenamefont {Felser},\ and\ \citenamefont {Parkin}}]{Nayak2016}%
  \BibitemOpen
  \bibfield  {author} {\bibinfo {author} {\bibfnamefont {A.~K.}\ \bibnamefont
  {Nayak}}, \bibinfo {author} {\bibfnamefont {J.~E.}\ \bibnamefont {Fischer}},
  \bibinfo {author} {\bibfnamefont {Y.}~\bibnamefont {Sun}}, \bibinfo {author}
  {\bibfnamefont {B.}~\bibnamefont {Yan}}, \bibinfo {author} {\bibfnamefont
  {J.}~\bibnamefont {Karel}}, \bibinfo {author} {\bibfnamefont {A.~C.}\
  \bibnamefont {Komarek}}, \bibinfo {author} {\bibfnamefont {C.}~\bibnamefont
  {Shekhar}}, \bibinfo {author} {\bibfnamefont {N.}~\bibnamefont {Kumar}},
  \bibinfo {author} {\bibfnamefont {W.}~\bibnamefont {Schnelle}}, \bibinfo
  {author} {\bibfnamefont {J.}~\bibnamefont {K{\"u}bler}}, \bibinfo {author}
  {\bibfnamefont {C.}~\bibnamefont {Felser}}, \ and\ \bibinfo {author}
  {\bibfnamefont {S.~S.~P.}\ \bibnamefont {Parkin}},\ }\href {\doibase
  10.1126/sciadv.1501870} {\bibfield  {journal} {\bibinfo  {journal} {Science
  Advances}\ }\textbf {\bibinfo {volume} {2}} (\bibinfo {year} {2016}),\
  10.1126/sciadv.1501870}\BibitemShut {NoStop}%
\bibitem [{\citenamefont {Ikhlas}\ \emph {et~al.}(2017)\citenamefont {Ikhlas},
  \citenamefont {Tomita}, \citenamefont {Koretsune}, \citenamefont {Suzuki},
  \citenamefont {Nishio-Hamane}, \citenamefont {Arita}, \citenamefont {Otani},\
  and\ \citenamefont {Nakatsuji}}]{Ikhlas2017}%
  \BibitemOpen
  \bibfield  {author} {\bibinfo {author} {\bibfnamefont {M.}~\bibnamefont
  {Ikhlas}}, \bibinfo {author} {\bibfnamefont {T.}~\bibnamefont {Tomita}},
  \bibinfo {author} {\bibfnamefont {T.}~\bibnamefont {Koretsune}}, \bibinfo
  {author} {\bibfnamefont {M.-T.}\ \bibnamefont {Suzuki}}, \bibinfo {author}
  {\bibfnamefont {D.}~\bibnamefont {Nishio-Hamane}}, \bibinfo {author}
  {\bibfnamefont {R.}~\bibnamefont {Arita}}, \bibinfo {author} {\bibfnamefont
  {Y.}~\bibnamefont {Otani}}, \ and\ \bibinfo {author} {\bibfnamefont
  {S.}~\bibnamefont {Nakatsuji}},\ }\href {http://dx.doi.org/10.1038/nphys4181}
  {\bibfield  {journal} {\bibinfo  {journal} {Nature Physics}\ }\textbf
  {\bibinfo {volume} {13}},\ \bibinfo {pages} {1085 EP } (\bibinfo {year}
  {2017})}\BibitemShut {NoStop}%
\bibitem [{\citenamefont {Chen}\ \emph {et~al.}(2014)\citenamefont {Chen},
  \citenamefont {Niu},\ and\ \citenamefont {MacDonald}}]{Chen2014}%
  \BibitemOpen
  \bibfield  {author} {\bibinfo {author} {\bibfnamefont {H.}~\bibnamefont
  {Chen}}, \bibinfo {author} {\bibfnamefont {Q.}~\bibnamefont {Niu}}, \ and\
  \bibinfo {author} {\bibfnamefont {A.~H.}\ \bibnamefont {MacDonald}},\ }\href
  {\doibase 10.1103/PhysRevLett.112.017205} {\bibfield  {journal} {\bibinfo
  {journal} {Phys. Rev. Lett.}\ }\textbf {\bibinfo {volume} {112}},\ \bibinfo
  {pages} {017205} (\bibinfo {year} {2014})}\BibitemShut {NoStop}%
\bibitem [{\citenamefont {Guo}\ and\ \citenamefont {Wang}(2017)}]{Guo2017}%
  \BibitemOpen
  \bibfield  {author} {\bibinfo {author} {\bibfnamefont {G.-Y.}\ \bibnamefont
  {Guo}}\ and\ \bibinfo {author} {\bibfnamefont {T.-C.}\ \bibnamefont {Wang}},\
  }\href {\doibase 10.1103/PhysRevB.96.224415} {\bibfield  {journal} {\bibinfo
  {journal} {Phys. Rev. B}\ }\textbf {\bibinfo {volume} {96}},\ \bibinfo
  {pages} {224415} (\bibinfo {year} {2017})}\BibitemShut {NoStop}%
\bibitem [{\citenamefont {Brown}\ \emph {et~al.}(1990)\citenamefont {Brown},
  \citenamefont {Nunez}, \citenamefont {Tasset}, \citenamefont {Forsyth},\ and\
  \citenamefont {Radhakrishna}}]{Brown1990}%
  \BibitemOpen
  \bibfield  {author} {\bibinfo {author} {\bibfnamefont {P.~J.}\ \bibnamefont
  {Brown}}, \bibinfo {author} {\bibfnamefont {V.}~\bibnamefont {Nunez}},
  \bibinfo {author} {\bibfnamefont {F.}~\bibnamefont {Tasset}}, \bibinfo
  {author} {\bibfnamefont {J.~B.}\ \bibnamefont {Forsyth}}, \ and\ \bibinfo
  {author} {\bibfnamefont {P.}~\bibnamefont {Radhakrishna}},\ }\href
  {http://stacks.iop.org/0953-8984/2/i=47/a=015} {\bibfield  {journal}
  {\bibinfo  {journal} {Journal of Physics: Condensed Matter}\ }\textbf
  {\bibinfo {volume} {2}},\ \bibinfo {pages} {9409} (\bibinfo {year}
  {1990})}\BibitemShut {NoStop}%
\bibitem [{\citenamefont {Feng}\ \emph {et~al.}(2006)\citenamefont {Feng},
  \citenamefont {Li}, \citenamefont {Ren}, \citenamefont {Li}, \citenamefont
  {Li}, \citenamefont {Li}, \citenamefont {Zhang},\ and\ \citenamefont
  {Zhang}}]{Feng2006}%
  \BibitemOpen
  \bibfield  {author} {\bibinfo {author} {\bibfnamefont {W.~J.}\ \bibnamefont
  {Feng}}, \bibinfo {author} {\bibfnamefont {D.}~\bibnamefont {Li}}, \bibinfo
  {author} {\bibfnamefont {W.~J.}\ \bibnamefont {Ren}}, \bibinfo {author}
  {\bibfnamefont {Y.~B.}\ \bibnamefont {Li}}, \bibinfo {author} {\bibfnamefont
  {W.~F.}\ \bibnamefont {Li}}, \bibinfo {author} {\bibfnamefont
  {J.}~\bibnamefont {Li}}, \bibinfo {author} {\bibfnamefont {Y.~Q.}\
  \bibnamefont {Zhang}}, \ and\ \bibinfo {author} {\bibfnamefont {Z.~D.}\
  \bibnamefont {Zhang}},\ }\href {\doibase 10.1103/PhysRevB.73.205105}
  {\bibfield  {journal} {\bibinfo  {journal} {Phys. Rev. B}\ }\textbf {\bibinfo
  {volume} {73}},\ \bibinfo {pages} {205105} (\bibinfo {year}
  {2006})}\BibitemShut {NoStop}%
\bibitem [{\citenamefont {Qian}\ \emph {et~al.}(2014)\citenamefont {Qian},
  \citenamefont {Nayak}, \citenamefont {Kreiner}, \citenamefont {Schnelle},\
  and\ \citenamefont {Felser}}]{Qian2014}%
  \BibitemOpen
  \bibfield  {author} {\bibinfo {author} {\bibfnamefont {J.~F.}\ \bibnamefont
  {Qian}}, \bibinfo {author} {\bibfnamefont {A.~K.}\ \bibnamefont {Nayak}},
  \bibinfo {author} {\bibfnamefont {G.}~\bibnamefont {Kreiner}}, \bibinfo
  {author} {\bibfnamefont {W.}~\bibnamefont {Schnelle}}, \ and\ \bibinfo
  {author} {\bibfnamefont {C.}~\bibnamefont {Felser}},\ }\href {\doibase
  10.1088/0022-3727/47/30/305001} {\bibfield  {journal} {\bibinfo  {journal}
  {Journal of Physics D: Applied Physics}\ }\textbf {\bibinfo {volume} {47}},\
  \bibinfo {pages} {305001} (\bibinfo {year} {2014})}\BibitemShut {NoStop}%
\bibitem [{\citenamefont {Yamada}\ \emph {et~al.}(1988)\citenamefont {Yamada},
  \citenamefont {Sakai}, \citenamefont {Mori},\ and\ \citenamefont
  {Ohoyama}}]{Yamada1988}%
  \BibitemOpen
  \bibfield  {author} {\bibinfo {author} {\bibfnamefont {N.}~\bibnamefont
  {Yamada}}, \bibinfo {author} {\bibfnamefont {H.}~\bibnamefont {Sakai}},
  \bibinfo {author} {\bibfnamefont {H.}~\bibnamefont {Mori}}, \ and\ \bibinfo
  {author} {\bibfnamefont {T.}~\bibnamefont {Ohoyama}},\ }\href {\doibase
  https://doi.org/10.1016/0378-4363(88)90258-6} {\bibfield  {journal} {\bibinfo
   {journal} {Physica B+C}\ }\textbf {\bibinfo {volume} {149}},\ \bibinfo
  {pages} {311 } (\bibinfo {year} {1988})}\BibitemShut {NoStop}%
\bibitem [{\citenamefont {Tomiyoshi}\ and\ \citenamefont
  {Yamaguchi}(1982)}]{Tomiyoshi1982}%
  \BibitemOpen
  \bibfield  {author} {\bibinfo {author} {\bibfnamefont {S.}~\bibnamefont
  {Tomiyoshi}}\ and\ \bibinfo {author} {\bibfnamefont {Y.}~\bibnamefont
  {Yamaguchi}},\ }\href {\doibase 10.1143/JPSJ.51.2478} {\bibfield  {journal}
  {\bibinfo  {journal} {Journal of the Physical Society of Japan}\ }\textbf
  {\bibinfo {volume} {51}},\ \bibinfo {pages} {2478} (\bibinfo {year}
  {1982})},\ \Eprint
  {http://arxiv.org/abs/https://doi.org/10.1143/JPSJ.51.2478}
  {https://doi.org/10.1143/JPSJ.51.2478} \BibitemShut {NoStop}%
\bibitem [{\citenamefont {Nagamiya}\ \emph {et~al.}(1982)\citenamefont
  {Nagamiya}, \citenamefont {Tomiyoshi},\ and\ \citenamefont
  {Yamaguchi}}]{Nagamiya1982}%
  \BibitemOpen
  \bibfield  {author} {\bibinfo {author} {\bibfnamefont {T.}~\bibnamefont
  {Nagamiya}}, \bibinfo {author} {\bibfnamefont {S.}~\bibnamefont {Tomiyoshi}},
  \ and\ \bibinfo {author} {\bibfnamefont {Y.}~\bibnamefont {Yamaguchi}},\
  }\href {\doibase https://doi.org/10.1016/0038-1098(82)90159-4} {\bibfield
  {journal} {\bibinfo  {journal} {Solid State Communications}\ }\textbf
  {\bibinfo {volume} {42}},\ \bibinfo {pages} {385 } (\bibinfo {year}
  {1982})}\BibitemShut {NoStop}%
\bibitem [{\citenamefont {Yang}\ \emph {et~al.}(2017)\citenamefont {Yang},
  \citenamefont {Sun}, \citenamefont {Zhang}, \citenamefont {Shi},
  \citenamefont {Parkin},\ and\ \citenamefont {Yan}}]{Yang2017}%
  \BibitemOpen
  \bibfield  {author} {\bibinfo {author} {\bibfnamefont {H.}~\bibnamefont
  {Yang}}, \bibinfo {author} {\bibfnamefont {Y.}~\bibnamefont {Sun}}, \bibinfo
  {author} {\bibfnamefont {Y.}~\bibnamefont {Zhang}}, \bibinfo {author}
  {\bibfnamefont {W.-J.}\ \bibnamefont {Shi}}, \bibinfo {author} {\bibfnamefont
  {S.~S.~P.}\ \bibnamefont {Parkin}}, \ and\ \bibinfo {author} {\bibfnamefont
  {B.}~\bibnamefont {Yan}},\ }\href
  {http://stacks.iop.org/1367-2630/19/i=1/a=015008} {\bibfield  {journal}
  {\bibinfo  {journal} {New Journal of Physics}\ }\textbf {\bibinfo {volume}
  {19}},\ \bibinfo {pages} {015008} (\bibinfo {year} {2017})}\BibitemShut
  {NoStop}%
\bibitem [{\citenamefont {Tomiyoshi}\ \emph {et~al.}(1983)\citenamefont
  {Tomiyoshi}, \citenamefont {Yamaguchi},\ and\ \citenamefont
  {Nagamiya}}]{Tomiyoshi1983}%
  \BibitemOpen
  \bibfield  {author} {\bibinfo {author} {\bibfnamefont {S.}~\bibnamefont
  {Tomiyoshi}}, \bibinfo {author} {\bibfnamefont {Y.}~\bibnamefont
  {Yamaguchi}}, \ and\ \bibinfo {author} {\bibfnamefont {T.}~\bibnamefont
  {Nagamiya}},\ }\href {\doibase https://doi.org/10.1016/0304-8853(83)90610-8}
  {\bibfield  {journal} {\bibinfo  {journal} {Journal of Magnetism and Magnetic
  Materials}\ }\textbf {\bibinfo {volume} {31-34}},\ \bibinfo {pages} {629 }
  (\bibinfo {year} {1983})}\BibitemShut {NoStop}%
\bibitem [{Note1()}]{Note1}%
  \BibitemOpen
  \bibinfo {note} {\label {Preprint}Upon finalizing this manuscript we became
  aware of the research of Xu \protect \textit {et al.} (arXiv:1812.04339). The
  Nernst data is in well agreement to our measurement, however, only one
  configuration of $S_{ij}$ was measured and the focus is lying on the
  relations between different transverse transport coefficients such as the
  Wiedemann-Franz law. The calculated SOC-induced gap that this work is showing
  does not influence our argument.}\BibitemShut {Stop}%
\bibitem [{Note2()}]{Note2}%
  \BibitemOpen
  \bibinfo {note} {For simplicity, a special index for highlighting those
  coefficients as anomalous is omitted in the following.}\BibitemShut {Stop}%
\bibitem [{\citenamefont {Meinero}\ \emph {et~al.}(2018)\citenamefont
  {Meinero}, \citenamefont {Caglieris}, \citenamefont {Lamura}, \citenamefont
  {Pallecchi}, \citenamefont {Jost}, \citenamefont {Zeitler}, \citenamefont
  {Ishida}, \citenamefont {Eisaki},\ and\ \citenamefont {Putti}}]{Meinero2018}%
  \BibitemOpen
  \bibfield  {author} {\bibinfo {author} {\bibfnamefont {M.}~\bibnamefont
  {Meinero}}, \bibinfo {author} {\bibfnamefont {F.}~\bibnamefont {Caglieris}},
  \bibinfo {author} {\bibfnamefont {G.}~\bibnamefont {Lamura}}, \bibinfo
  {author} {\bibfnamefont {I.}~\bibnamefont {Pallecchi}}, \bibinfo {author}
  {\bibfnamefont {A.}~\bibnamefont {Jost}}, \bibinfo {author} {\bibfnamefont
  {U.}~\bibnamefont {Zeitler}}, \bibinfo {author} {\bibfnamefont
  {S.}~\bibnamefont {Ishida}}, \bibinfo {author} {\bibfnamefont
  {H.}~\bibnamefont {Eisaki}}, \ and\ \bibinfo {author} {\bibfnamefont
  {M.}~\bibnamefont {Putti}},\ }\href {\doibase 10.1103/PhysRevB.98.155116}
  {\bibfield  {journal} {\bibinfo  {journal} {Phys. Rev. B}\ }\textbf {\bibinfo
  {volume} {98}},\ \bibinfo {pages} {155116} (\bibinfo {year}
  {2018})}\BibitemShut {NoStop}%
\bibitem [{\citenamefont {Xu}\ \emph {et~al.}(2018{\natexlab{a}})\citenamefont
  {Xu}, \citenamefont {Li}, \citenamefont {Lu}, \citenamefont {Collignon},
  \citenamefont {Fu}, \citenamefont {Fauqué}, \citenamefont {Yan},
  \citenamefont {Zhu},\ and\ \citenamefont {Behnia}}]{Xu2019}%
  \BibitemOpen
  \bibfield  {author} {\bibinfo {author} {\bibfnamefont {L.}~\bibnamefont
  {Xu}}, \bibinfo {author} {\bibfnamefont {X.}~\bibnamefont {Li}}, \bibinfo
  {author} {\bibfnamefont {X.}~\bibnamefont {Lu}}, \bibinfo {author}
  {\bibfnamefont {C.}~\bibnamefont {Collignon}}, \bibinfo {author}
  {\bibfnamefont {H.}~\bibnamefont {Fu}}, \bibinfo {author} {\bibfnamefont
  {B.}~\bibnamefont {Fauqué}}, \bibinfo {author} {\bibfnamefont
  {B.}~\bibnamefont {Yan}}, \bibinfo {author} {\bibfnamefont {Z.}~\bibnamefont
  {Zhu}}, \ and\ \bibinfo {author} {\bibfnamefont {K.}~\bibnamefont {Behnia}},\
  }\href@noop {} {\  (\bibinfo {year} {2018}{\natexlab{a}})},\ \Eprint
  {http://arxiv.org/abs/arXiv:1812.04339} {arXiv:1812.04339} \BibitemShut
  {NoStop}%
\bibitem [{\citenamefont {Sharma}\ \emph {et~al.}(2016)\citenamefont {Sharma},
  \citenamefont {Goswami},\ and\ \citenamefont {Tewari}}]{Sharma2016}%
  \BibitemOpen
  \bibfield  {author} {\bibinfo {author} {\bibfnamefont {G.}~\bibnamefont
  {Sharma}}, \bibinfo {author} {\bibfnamefont {P.}~\bibnamefont {Goswami}}, \
  and\ \bibinfo {author} {\bibfnamefont {S.}~\bibnamefont {Tewari}},\ }\href
  {\doibase 10.1103/PhysRevB.93.035116} {\bibfield  {journal} {\bibinfo
  {journal} {Phys. Rev. B}\ }\textbf {\bibinfo {volume} {93}},\ \bibinfo
  {pages} {035116} (\bibinfo {year} {2016})}\BibitemShut {NoStop}%
\bibitem [{\citenamefont {Kubo}(1957)}]{Kubo1957}%
  \BibitemOpen
  \bibfield  {author} {\bibinfo {author} {\bibfnamefont {R.}~\bibnamefont
  {Kubo}},\ }\href {\doibase 10.1143/JPSJ.12.570} {\bibfield  {journal}
  {\bibinfo  {journal} {Journal of the Physical Society of Japan}\ }\textbf
  {\bibinfo {volume} {12}},\ \bibinfo {pages} {570} (\bibinfo {year} {1957})},\
  \Eprint {http://arxiv.org/abs/https://doi.org/10.1143/JPSJ.12.570}
  {https://doi.org/10.1143/JPSJ.12.570} \BibitemShut {NoStop}%
\bibitem [{\citenamefont {Xu}\ \emph {et~al.}(2018{\natexlab{b}})\citenamefont
  {Xu}, \citenamefont {Li}, \citenamefont {Lu}, \citenamefont {Collignon},
  \citenamefont {Fu}, \citenamefont {Fauque}, \citenamefont {Yan},
  \citenamefont {Zhu},\ and\ \citenamefont {Behnia}}]{Xu2018}%
  \BibitemOpen
  \bibfield  {author} {\bibinfo {author} {\bibfnamefont {L.}~\bibnamefont
  {Xu}}, \bibinfo {author} {\bibfnamefont {X.}~\bibnamefont {Li}}, \bibinfo
  {author} {\bibfnamefont {X.}~\bibnamefont {Lu}}, \bibinfo {author}
  {\bibfnamefont {C.}~\bibnamefont {Collignon}}, \bibinfo {author}
  {\bibfnamefont {H.}~\bibnamefont {Fu}}, \bibinfo {author} {\bibfnamefont
  {B.}~\bibnamefont {Fauque}}, \bibinfo {author} {\bibfnamefont
  {B.}~\bibnamefont {Yan}}, \bibinfo {author} {\bibfnamefont {Z.}~\bibnamefont
  {Zhu}}, \ and\ \bibinfo {author} {\bibfnamefont {K.}~\bibnamefont {Behnia}},\
  }\href@noop {} {\bibfield  {journal} {\bibinfo  {journal} {arXiv:1812.04339}\
  } (\bibinfo {year} {2018}{\natexlab{b}})}\BibitemShut {NoStop}%
\bibitem [{\citenamefont {Nayak}\ \emph {et~al.}(2015)\citenamefont {Nayak},
  \citenamefont {Nicklas}, \citenamefont {Chadov}, \citenamefont {Khuntia},
  \citenamefont {Shekhar}, \citenamefont {Kalache}, \citenamefont {Baenitz},
  \citenamefont {Skourski}, \citenamefont {Guduru}, \citenamefont {Puri},
  \citenamefont {Zeitler}, \citenamefont {Coey},\ and\ \citenamefont
  {Felser}}]{Nayak2015}%
  \BibitemOpen
  \bibfield  {author} {\bibinfo {author} {\bibfnamefont {A.~K.}\ \bibnamefont
  {Nayak}}, \bibinfo {author} {\bibfnamefont {M.}~\bibnamefont {Nicklas}},
  \bibinfo {author} {\bibfnamefont {S.}~\bibnamefont {Chadov}}, \bibinfo
  {author} {\bibfnamefont {P.}~\bibnamefont {Khuntia}}, \bibinfo {author}
  {\bibfnamefont {C.}~\bibnamefont {Shekhar}}, \bibinfo {author} {\bibfnamefont
  {A.}~\bibnamefont {Kalache}}, \bibinfo {author} {\bibfnamefont
  {M.}~\bibnamefont {Baenitz}}, \bibinfo {author} {\bibfnamefont
  {Y.}~\bibnamefont {Skourski}}, \bibinfo {author} {\bibfnamefont {V.~K.}\
  \bibnamefont {Guduru}}, \bibinfo {author} {\bibfnamefont {A.}~\bibnamefont
  {Puri}}, \bibinfo {author} {\bibfnamefont {U.}~\bibnamefont {Zeitler}},
  \bibinfo {author} {\bibfnamefont {J.~M.~D.}\ \bibnamefont {Coey}}, \ and\
  \bibinfo {author} {\bibfnamefont {C.}~\bibnamefont {Felser}},\ }\href
  {http://dx.doi.org/10.1038/nmat4248} {\bibfield  {journal} {\bibinfo
  {journal} {Nature Materials}\ }\textbf {\bibinfo {volume} {14}},\ \bibinfo
  {pages} {679 EP } (\bibinfo {year} {2015})}\BibitemShut {NoStop}%
\bibitem [{\citenamefont {Park}\ \emph {et~al.}(2011)\citenamefont {Park},
  \citenamefont {Wunderlich}, \citenamefont {Mart{\'\i}}, \citenamefont
  {Hol{\'y}}, \citenamefont {Kurosaki}, \citenamefont {Yamada}, \citenamefont
  {Yamamoto}, \citenamefont {Nishide}, \citenamefont {Hayakawa}, \citenamefont
  {Takahashi}, \citenamefont {Shick},\ and\ \citenamefont
  {Jungwirth}}]{Park2011}%
  \BibitemOpen
  \bibfield  {author} {\bibinfo {author} {\bibfnamefont {B.~G.}\ \bibnamefont
  {Park}}, \bibinfo {author} {\bibfnamefont {J.}~\bibnamefont {Wunderlich}},
  \bibinfo {author} {\bibfnamefont {X.}~\bibnamefont {Mart{\'\i}}}, \bibinfo
  {author} {\bibfnamefont {V.}~\bibnamefont {Hol{\'y}}}, \bibinfo {author}
  {\bibfnamefont {Y.}~\bibnamefont {Kurosaki}}, \bibinfo {author}
  {\bibfnamefont {M.}~\bibnamefont {Yamada}}, \bibinfo {author} {\bibfnamefont
  {H.}~\bibnamefont {Yamamoto}}, \bibinfo {author} {\bibfnamefont
  {A.}~\bibnamefont {Nishide}}, \bibinfo {author} {\bibfnamefont
  {J.}~\bibnamefont {Hayakawa}}, \bibinfo {author} {\bibfnamefont
  {H.}~\bibnamefont {Takahashi}}, \bibinfo {author} {\bibfnamefont {A.~B.}\
  \bibnamefont {Shick}}, \ and\ \bibinfo {author} {\bibfnamefont
  {T.}~\bibnamefont {Jungwirth}},\ }\href {http://dx.doi.org/10.1038/nmat2983}
  {\bibfield  {journal} {\bibinfo  {journal} {Nature Materials}\ }\textbf
  {\bibinfo {volume} {10}},\ \bibinfo {pages} {347 EP } (\bibinfo {year}
  {2011})}\BibitemShut {NoStop}%
\bibitem [{\citenamefont {Marti}\ \emph {et~al.}(2014)\citenamefont {Marti},
  \citenamefont {Fina}, \citenamefont {Frontera}, \citenamefont {Liu},
  \citenamefont {Wadley}, \citenamefont {He}, \citenamefont {Paull},
  \citenamefont {Clarkson}, \citenamefont {Kudrnovsk{\'y}}, \citenamefont
  {Turek}, \citenamefont {Kune{\v s}}, \citenamefont {Yi}, \citenamefont {Chu},
  \citenamefont {Nelson}, \citenamefont {You}, \citenamefont {Arenholz},
  \citenamefont {Salahuddin}, \citenamefont {Fontcuberta}, \citenamefont
  {Jungwirth},\ and\ \citenamefont {Ramesh}}]{Marti2014}%
  \BibitemOpen
  \bibfield  {author} {\bibinfo {author} {\bibfnamefont {X.}~\bibnamefont
  {Marti}}, \bibinfo {author} {\bibfnamefont {I.}~\bibnamefont {Fina}},
  \bibinfo {author} {\bibfnamefont {C.}~\bibnamefont {Frontera}}, \bibinfo
  {author} {\bibfnamefont {J.}~\bibnamefont {Liu}}, \bibinfo {author}
  {\bibfnamefont {P.}~\bibnamefont {Wadley}}, \bibinfo {author} {\bibfnamefont
  {Q.}~\bibnamefont {He}}, \bibinfo {author} {\bibfnamefont {R.~J.}\
  \bibnamefont {Paull}}, \bibinfo {author} {\bibfnamefont {J.~D.}\ \bibnamefont
  {Clarkson}}, \bibinfo {author} {\bibfnamefont {J.}~\bibnamefont
  {Kudrnovsk{\'y}}}, \bibinfo {author} {\bibfnamefont {I.}~\bibnamefont
  {Turek}}, \bibinfo {author} {\bibfnamefont {J.}~\bibnamefont {Kune{\v s}}},
  \bibinfo {author} {\bibfnamefont {D.}~\bibnamefont {Yi}}, \bibinfo {author}
  {\bibfnamefont {J.-H.}\ \bibnamefont {Chu}}, \bibinfo {author} {\bibfnamefont
  {C.~T.}\ \bibnamefont {Nelson}}, \bibinfo {author} {\bibfnamefont
  {L.}~\bibnamefont {You}}, \bibinfo {author} {\bibfnamefont {E.}~\bibnamefont
  {Arenholz}}, \bibinfo {author} {\bibfnamefont {S.}~\bibnamefont
  {Salahuddin}}, \bibinfo {author} {\bibfnamefont {J.}~\bibnamefont
  {Fontcuberta}}, \bibinfo {author} {\bibfnamefont {T.}~\bibnamefont
  {Jungwirth}}, \ and\ \bibinfo {author} {\bibfnamefont {R.}~\bibnamefont
  {Ramesh}},\ }\href {http://dx.doi.org/10.1038/nmat3861} {\bibfield  {journal}
  {\bibinfo  {journal} {Nature Materials}\ }\textbf {\bibinfo {volume} {13}},\
  \bibinfo {pages} {367 EP } (\bibinfo {year} {2014})}\BibitemShut {NoStop}%
\bibitem [{\citenamefont {Mizukami}\ \emph {et~al.}(2013)\citenamefont
  {Mizukami}, \citenamefont {Sakuma}, \citenamefont {Sugihara}, \citenamefont
  {Kubota}, \citenamefont {Kondo}, \citenamefont {Tsuchiura},\ and\
  \citenamefont {Miyazaki}}]{Mizukami2013}%
  \BibitemOpen
  \bibfield  {author} {\bibinfo {author} {\bibfnamefont {S.}~\bibnamefont
  {Mizukami}}, \bibinfo {author} {\bibfnamefont {A.}~\bibnamefont {Sakuma}},
  \bibinfo {author} {\bibfnamefont {A.}~\bibnamefont {Sugihara}}, \bibinfo
  {author} {\bibfnamefont {T.}~\bibnamefont {Kubota}}, \bibinfo {author}
  {\bibfnamefont {Y.}~\bibnamefont {Kondo}}, \bibinfo {author} {\bibfnamefont
  {H.}~\bibnamefont {Tsuchiura}}, \ and\ \bibinfo {author} {\bibfnamefont
  {T.}~\bibnamefont {Miyazaki}},\ }\href
  {http://stacks.iop.org/1882-0786/6/i=12/a=123002} {\bibfield  {journal}
  {\bibinfo  {journal} {Applied Physics Express}\ }\textbf {\bibinfo {volume}
  {6}},\ \bibinfo {pages} {123002} (\bibinfo {year} {2013})}\BibitemShut
  {NoStop}%
\end{thebibliography}
\end{document}